\documentclass[default,iicol]{sn-jnl}

\jyear{2021}%

\theoremstyle{thmstyleone}%
\theoremstyle{thmstyletwo}%

\theoremstyle{thmstylethree}%

\begin{document}

\title[Rolling relaxation controls friction weakening]{Frictional weakening of a granular sheared layer due to viscous rolling revealed by Discrete Element Modeling}


\author*[1,2]{\fnm{Alexandre} \sur{Sac-Morane}}\email{alexandre.sac-morane@uclouvain.be}

\author[1]{\fnm{Manolis} \sur{Veveakis}}\email{manolis.veveakis@duke.edu}
\equalcont{These authors contributed equally to this work.}

\author[2]{\fnm{Hadrien} \sur{Rattez}}\email{hadrien.rattez@uclouvain.be}
\equalcont{These authors contributed equally to this work.}

\affil*[1]{\orgdiv{Multiphysics Geomechanics Lab}, \orgname{Duke University}, \orgaddress{\city{Durham}, \postcode{27708}, \state{NC}, \country{USA}}}

\affil[2]{\orgdiv{Institute of Mechanics, Materials and Civil Engineering}, \orgname{UCLouvain}, \orgaddress{\city{Louvain-la-Neuve}, \postcode{1348}, \country{Belgium}}}


\abstract{Considering a 3D sheared granular layer through a discrete element modeling, it is well known the rolling resistance influences the macro friction coefficient. Even if the rolling resistance role has been deeply investigated previously because it is commonly used to represent the shape and the roughness of the grains, the rolling viscous damping coefficient is still not studied. This parameter is rarely used or only to dissipate the energy and to converge numerically. This paper revisits the physical role of those coefficients with a parametric study of the rolling friction and the rolling damping at different shear speeds and different confinement pressures. It has been observed the damping coefficient induces a frictional weakening. Hence, competition between the rolling resistance and the rolling damping occurs. Angular resistance aims to avoid grains rolling, decreasing the difference between the angular velocities of grains. Whereas, angular damping acts in the opposite, avoiding a change in the difference between the angular velocities of grains. In consequence, grains stay rolling and the sample toughness decreases. This effect must be considered to not overestimate the frictional response of a granular layer.}

\keywords{Discrete element method, Rolling parameter, Sheared layer friction, Granular materials}

\maketitle

\section{Introduction}\label{sec1}

Accurately measuring or calculating the frictional strength of granular sheared layers is of paramount importance across all fields of granular media-related sciences, including earthquakes and fault mechanics \cite{MYERS2004947, Poulet2014}, landslides \cite{Segui2020}, and debris flows \cite{https://agupubs.onlinelibrary.wiley.com/doi/pdf/10.1029/97RG00426} to name but a few. It is very well accepted nowadays that the calculation of a macroscopic property like the frictional coefficient of granular media is the result of grain-to-grain interactions at the micro-scale \cite{SulemVardoulakis1995, Sulemetal2011}. Therefore, in order to accurate capture those effects and homogenize them to the friction coefficient of a layer, higher order analytical and numerical approaches need to be considered \cite{Rattez2018a,Rattez2018b,Rattez2018c}.

One of the most well accepted approaches in direct modeling of granular media is the Discrete Element Method \cite{https://agupubs.onlinelibrary.wiley.com/doi/10.1029/2022JB025209?af=R}, which has been designed to consider those interactions between grains \cite{Burman1980}. The method started with a simple linear contact law\cite{PhysRevE.89.042210,PhysRevE.92.022202}, but since then contact laws have been modified by \cite{OSullivan2011}: (i) considering the grain crushing \cite{HANLEY20151100,Zhang2021}, (ii) investigating the effect of the pressure solution \cite{Rutter1976,LEHNER1995153,VandenEnde2018}, (iii) exploring the effect of the healing \cite{Abe2002, Morgan2004}, (iv) appreciating the influence of the cohesion in the matter \cite{Potyondy2004,Li2017,Casas2020} or (v) the cohesion induced by the pore fluid \cite{Soulie2006,Dorostkar2018} among others. Also, it allows some to focus on the temperature influence, identifying the pressurization of the pore fluid \cite{doi:10.1680/geot.2002.52.3.157,Rice2006} and grain melting \cite{Gan2012,Mollon2021} as the main phenomena driving the evolution of the frictional strength of a fault zone during large crustal events.

The present work is motivated by previous experiments made on antigorite \cite{Idrissi2020}, and the main goal is to assess the influence on global behavior of contact laws and parameters values. Even though different relevant outputs for dense granular flows are reviewed by the French research group \textit{Groupement de Recherche Milieux Divisés} (GDR MiDi) \cite{Midi2004}, we are focused in this paper on the macro friction coefficient at steady state. As such, we introduce and explore the influence of the rolling resistance between grains to the macroscopic strength of a granular sheared layer. Experimental results \cite{doi:10.1680/geot.2004.54.8.539,ODA1982269}, and numerical ones \cite{Zhou1999,Alonso-Marroquin2006,Papanicolopulos2011} have highlighted that grain rolling has a real impact on the sample behavior with many rolling models being formulated since \cite{Ai2011,Zhao2016}. In the literature the  elastic-plastic spring dashpot model is identified as the benchmark for this response \cite{IwashitaK.Oda1998,Iwashita2000} and has been extended to conclude that: (i) rolling helps the formation of shear bands and decrease the sample strength \cite{Murakami1997,Zhang2013,Tang2016,Nho2021}; (ii) the stress-dilatancy curves are modified \cite{Estrada2008,Yang2017,Liu2018,Barnett2020} when accounting for the rolling resistance coming from intragranular friction \cite{GODET1984437,COLAS2013192} and roughness \cite{https://doi.org/10.1002/(SICI)1096-9853(199905)23:6<531::AID-NAG980>3.0.CO;2-V,Kozicki2011,Mollon2020}. 

However, the computational cost of these approaches is not negligible, and especially if grains clusters \cite{Garcia2009}, superquadric particles \cite{Podlozhnyuk2018} or even polyhedral shapes \cite{Cundall1988, Nezami2004, AlonsoMarroquin2009} are assumed to approximate the shape, those simulations become quickly computationally costly. Because of this fact, geometric laws for the rolling friction have been developed \cite{Wensrich2012,Rorato2021} allowing for simulations to keep using round particles with a rolling resistance stemming from an equivalent shape. However, the introduced angular damping influence is not well constrained, hence being neglected in most of the DEM simulations only used for stability reasons \cite{Ai2011,IwashitaK.Oda1998} rather than for physical robustness \cite{Jiang2005} In this work we revisit the physical role of the rolling resistance in a granular sheared layer and perform a parametric study over the rolling friction and the rolling viscous damping coefficients to understand better their influence on the macroscopic friction coefficient of a granular sheared layer. 

\section{Theory and formulation}
The Discrete Element Model (DEM) is an approach developed by Cundall \& Strack \cite{Burman1980} to simulate granular materials at the particles level. The foundation of this method is to consider inside the material the individual particles and their interactions explicitly. Newton's laws (linear and angular momentum) are used to compute the motion of the grains, as follows:

\begin{equation}
    m\frac{\partial v_i}{\partial t} = m\times g_i + \sum f_i
    \label{Newton Law 1}
\end{equation}

\begin{equation}
    I\times\rho\frac{\partial\omega_i}{\partial t}=\sum \epsilon_{ijk}f_jR_k + \sum M_i
    \label{Newton Law 2}
\end{equation}

where $m$ is the particle mass, $v$ the particle velocity, $g$ the gravity acceleration, $f$ the contact forces, $I$ the moment of inertia, $\rho$ the particle density, $\omega$ the angular velocity, $\epsilon_{ijk}$ the Levi-Civita symbol, $R$ the radius, $M$ the contact moments.

Considering two particles with radii $R^1$ and $R^2$, the interaction between particles is computed only if the distance $\delta$ between grains satisfies the following inequality:
\begin{equation}
    \delta<R^1+R^2
    \label{overlap condition}
\end{equation}

Once contact is detected between grains $1$ and $2$, interactions (force and moment) are computed from relative motions $\Delta u$ and $\Delta \omega$ as:

\begin{equation}
    \Delta u_i = u^1_i - u^2_i + \epsilon_{ijk}\left(R^1_j\theta^1_k - R^2_j\theta^2_k\right)
    \label{Delta u}
\end{equation}

\begin{equation}
    \Delta \omega_i=\omega^1_i-\omega^2_i
    \label{Delta omega}
\end{equation}

where $u$ is the particle displacement, $\theta$ the angular displacement and $\omega$ the angular velocity of the grain.

\begin{figure}[h]
    \centering
\includegraphics[width=0.9\linewidth]{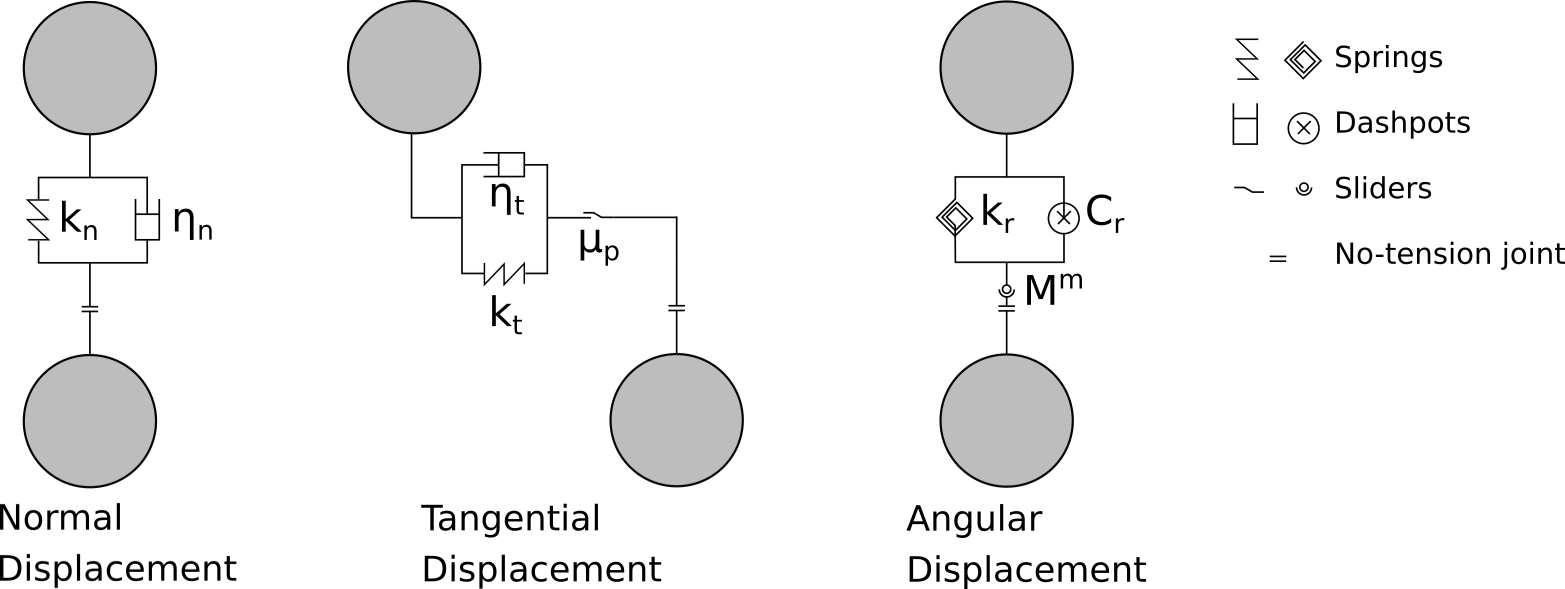}
    \caption{The contact between two particles obeys to normal, tangential and rolling elastic-plastic spring-dashpot laws.}
    \label{DEM Scheme}
\end{figure}

The contact models between cohesionless particles obey the Hertz contact theory \cite{Johnson1985}. Normal, tangential and angular models are shown at figure \ref{DEM Scheme} and described in the following. 

As the contact can happen between particles with different properties, some equivalent parameters need to be defined. The equivalent radius $R^*$ and equivalent mass $m^*$ are defined at equations \ref{Equivalent radius} and \ref{Equivalent mass} with an harmonic mean. 

\begin{equation}
\frac{1}{R^*} = \frac{1}{R^1} + \frac{1}{R^2}
    \label{Equivalent radius}
\end{equation}
\begin{equation}
\frac{1}{m^*} = \frac{1}{m^1} + \frac{1}{m^2}
    \label{Equivalent mass}
\end{equation}

The equivalent Young modulus $Y^*$ and shear modulus $G^*$ are defined at equations \ref{Equivalent Young modulus} and \ref{Equivalent shear modulus} with an harmonic mean adjusted by Poisson's ratio.

\begin{equation}
\begin{split}
\frac{1}{Y^*} = \frac{1-\nu^{1^2}}{Y^1} + \frac{1-\nu^{2^2}}{Y^2} \\
\Longleftrightarrow Y^*=\frac{Y}{2(1-\nu^2)}
\end{split}
    \label{Equivalent Young modulus}
\end{equation}
\begin{equation}
\begin{split}
\frac{1}{G^*} = \frac{2(2-\nu^1)(1+\nu^1)}{Y^1} + \frac{2(2-\nu^2)(1+\nu^2)}{Y^2} \\
\Longleftrightarrow G^* = \frac{Y}{4(2-\nu)(1+\nu)}
\end{split}
    \label{Equivalent shear modulus}
\end{equation}

The equivalent moment of inertia $I^*$ is defined at equation \ref{Equivalent Moment Inertia} with an harmonic mean of the different moments of inertia displaced at the contact point.

\begin{equation}
\frac{1}{I^*} = \frac{1}{I^1 + m^1R^{1^2}} +\frac{1}{I^2 + m^2R^{2^2}}
    \label{Equivalent Moment Inertia}
\end{equation}

\vskip\baselineskip

\emph{Normal model}\\
The normal force is formulated as:
\begin{equation}
f_n = k_n\Delta_n - \gamma_nv_n
\label{Force Normal}
\end{equation}

The reaction is divided into a spring part and a damping part, with the normal stiffness $k_n$ formulated as: 
\begin{equation}
k_n = \frac{4}{3} Y^* \sqrt{R^* \Delta_n}
\label{Spring Normal}
\end{equation}
Following the Hertz contact theory, this parameter depends mainly on the normal overlap $\Delta_n$. Thus, the normal force is not linear with respect to the overlap. The normal stiffness depends also on the equivalent Young modulus $Y^*$ and the equivalent radius $R^*$ defined before. The normal damping $\gamma_n$ is null in this paper because the restitution coefficient $e$ is taken at the value $1$. This choice have been made to focus on the influence of the rolling damping.

\vskip\baselineskip

\emph{Tangential model}\\
The tangential force is formulated to verify the Coulomb friction law defined on the friction coefficient between particle $\mu_p$ and the normal force $f_n$:

\begin{equation}
f_t = k_t\Delta_t - \gamma_tv_t \leq \mu_p f_n
\label{Force Tangential}
\end{equation}

The reaction is divided into a spring part and a damping part. The tangential stiffness $k_t$ is formulated as:
\begin{equation}
k_t = 8  G^* \sqrt{R^*\Delta_n}
\label{Spring Tangential}
\end{equation}
Following the Hertz contact theory, this parameter depends mainly on the normal overlap $\Delta_n$. Thus, the tangential force is not linear with respect to the overlap. The tangential stiffness depends also on the equivalent shear modulus $G^*$ and the equivalent radius $R^*$ defined before. The tangential damping $\gamma_t$ is also null in this paper because the restitution coefficient $e$ is taken at the value $1$. 

\vskip\baselineskip

\emph{Angular model}\\
A lot of angular models could be applied but an elastic-plastic spring-dashpot model is used because it is the most accurate choice. Hence, the model allows energy dissipation during relative rotation and provides packing support for static system, two main functions to verify in a particulate system \cite{Ai2011}.
The reaction moment $M$ is formulated as:
\begin{equation}
M = M^k + M^d
\label{Rolling Model}
\end{equation}

This reaction is divided into a spring part $M^k$ and a damping part $M^d$ defined at equations \ref{Rolling Spring Model} and \ref{Rolling Damping Model}: 

\begin{equation} 
M^k_{t+\Delta t} = M^k_t -k_r \Delta \theta  \leq M^m
\label{Rolling Spring Model}
\end{equation}

The incremental angle $\Delta \theta$ is obtained by a time integration of the angular velocity $\Delta \omega \times dt$. The angular stiffness $k_r$ is formulated at equation \ref{Rolling Stiffness} by considering a continuously distributed system of normal and tangential spring at the interface \cite{Ai2011}, \cite{Jiang2005}.

\begin{equation}
    k_r = 2,25k_n\mu_r^2R^{*2}
    \label{Rolling Stiffness}
\end{equation}

The rolling friction coefficient $\mu_r$ is introduced. This variable is a dimensionless parameter defined as \cite{Ai2011}:

\begin{equation}
    \mu_r = tan(\beta)
    \label{Rolling friction coefficient}
\end{equation}

\begin{figure}[h]
    \centering
    \includegraphics[width=0.4\linewidth]{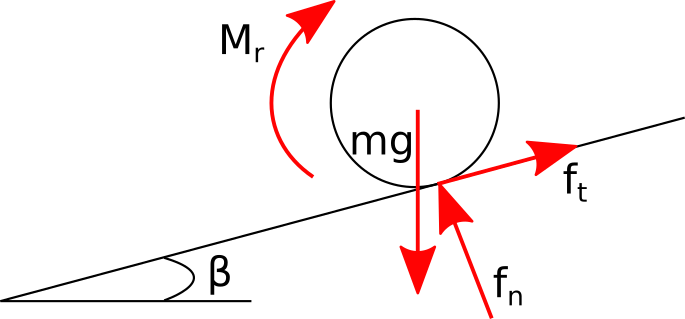}
    \caption{Definition of the rolling resistance coefficient $\mu_r = tan(\beta)$.}
    \label{RollingResistanceAngle}
\end{figure}

The angle $\beta$ represents the maximum angle of a slope on which the rolling resistance moment counterbalances the moment due to gravity on the grain, see figure \ref{RollingResistanceAngle}. The influence of $\mu_r$ is investigated in this paper.

\begin{equation}
    M^d_{t+\Delta t} =\left\lbrace
\begin{array}{ll} 
-C_r\Delta\omega & \text{ if } M^k_{t+\Delta t}<M^m\\
0 & \text{ if }  M^k_{t+\Delta t}=M^m\\
\end{array}\right.
\label{Rolling Damping Model}
\end{equation}

It appears the damping part $M^d$ is defined with a rolling viscous damping parameter $C_r$ formulated at equation \ref{Rolling Damping Coefficient}.

\begin{equation}
C_r = 2\eta_r\sqrt{I_rk_r}
\label{Rolling Damping Coefficient}
\end{equation}

The rolling viscous damping coefficient $\eta_r$ is introduced. This variable is a dimensionless parameter and its influence is investigated in this paper. Moreover, it appears the rolling viscous damping parameter depends on the rolling stiffness $k_r$ and so on the rolling friction coefficient $\mu_r$.

As described by equations \ref{Rolling Spring Model} and \ref{Rolling Damping Model}, the spring and damping parts are restricted by a plastic behavior, the rolling of particles. The rolling starts when the spring part reaches the plastic limit $M^m$ defined at equation \ref{Angular Plastic Limit}.

\begin{equation} 
M^m=\mu_rR^*f_n
\label{Angular Plastic Limit}
\end{equation}

This limit depends on the equivalent radius, the rolling friction coefficient and the normal force. Once rolling occurs, the reaction from the angular spring takes the value of $M^m$ and the damping element is deleted. 

\section{Numerical model}

\begin{figure}[h]
\centering
\includegraphics[width=0.6\linewidth]{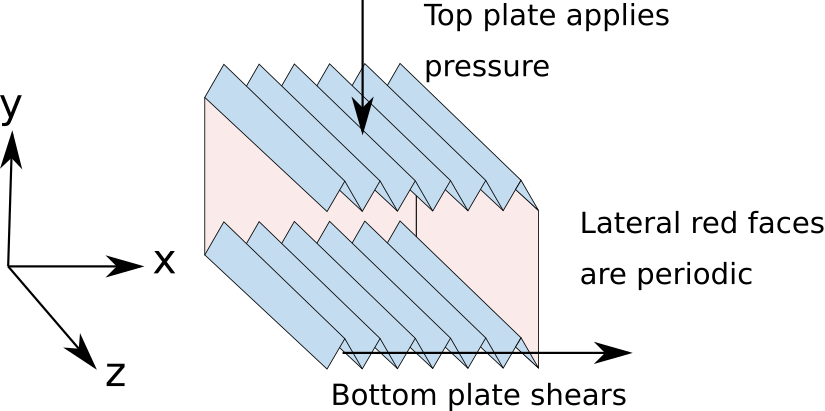}
\caption{The simulation box with triangle plates and periodic faces.}
\label{Schema Box}
\end{figure}

\begin{figure}[h]
\centering
\includegraphics[width=0.9\linewidth]{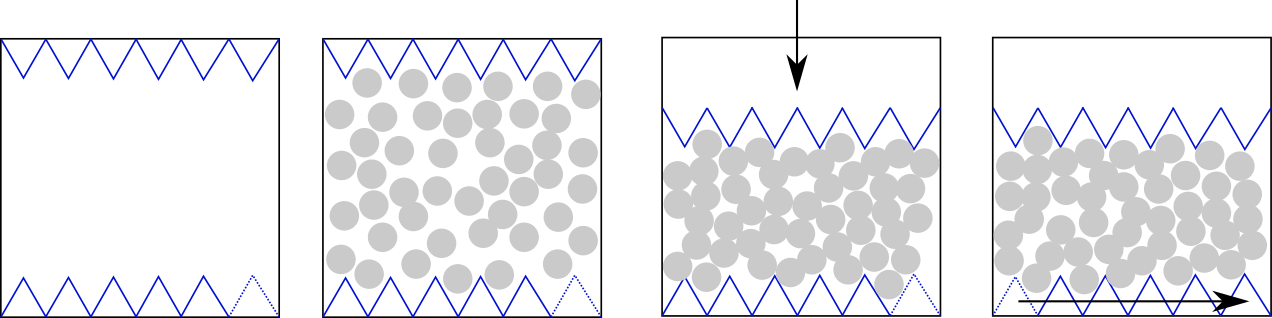}
\caption{The simulation is in multiple steps : creation of the box and particles, application of the normal force and shearing.}
\label{Simulation Steps}
\end{figure}

The simulation setup is illustrated at figure \ref{Schema Box}. The box is a $0,004\,m\times 0,006\,m\times 0,0024\,m$ region. Faces x and z are under periodic conditions. The size of domain has been chosen to respect a sufficient number of grain over the different axis. On the axis x, the shearing direction, there are $l_x/d_{50} = 4/0.26 = 15$ particles. On the axis z, the minor direction, there are $l_z/d_{50} = 2.4/0.26 = 10$ particles. On the axis y, the size allows the particles generation shown at figure \ref{Simulation Steps}. Sizes have been minimized to reduce the number of grains and so the computational cost, main trouble with DEM simulations. The gravity is not considered because its effect stays negligible under the vertical pressure applied.

 The simulation, made by the open-source software LIGGGHTS \cite{Kloss2012}, is in several steps illustrated at figure \ref{Simulation Steps}:
\begin{enumerate}
\item The box, bottom and top triangle plates are created. The triangle pattern represents the roughness of the plates with a geometry similar to experimental tests \cite{Marone2005,Koval2011}. The specific size of the triangle is defined to be $1.5$ times the largest particle diameter.
\item 2500 particles are generated following the distribution presented in table \ref{Distribution} equivalent to the one used in \cite{Morgan2004}. This number of particles allows to get $17$ particles on the height where the residual shear band sizes for different distributions was assumed to be between $9\times d_{50}$ and $16\times d_{50}$ \cite{Rattez2020}.
\item Top plate applies vertical stress of 10 MPa by moving following the y axis. This plate is free to move vertically to verify this confining and allow volume change. The value of the vertical stress has been chosen from previous experiments and numerical simulations \cite{Dieterich1979,Morrow1989,Ferdowsi2020}.
\item The sample is sheared by moving the bottom plate at the speed of 100 $\mu$m/s until 100\% strain. This step is then repeated at the speed of 300 and 1000 $\mu$m/s. Those velocities have been chosen from previous numerical simulations \cite{Ferdowsi2020} and in-situ estimations \cite{Beroza1990} to minimize the computational cost but still to represent real sheared layers. 
\end{enumerate}

\begin{table}[h]
\centering
\begin{tabular}{|l|l|l|}
\hline
Radius&Percentage&Number of particles\\
\hline
R1 $= 0,2\,mm$ & $14\%$ & 2500 \\
R2 $= 0,15\,mm$ & $29\%$ &\\
R3 $= 0,1\,mm$ & $57\%$ & \\
\hline
\end{tabular}
\caption{Distribution used described by discrete radius, percentage of the mass and total number of grains.}
\label{Distribution}
\end{table}

\begin{table}[h]
\begin{tabular}{|p{0.3\linewidth}|p{0.12\linewidth}|p{0.45\linewidth}|}
\hline
Variable&Short Name&Value\\
\hline
\multicolumn{3}{|c|}{Simulation variables}\\
\hline
Time step&$dt$&$1,5e^{-6}$ s\\
Height of the sample&$h$&$0,005$ m\\
Shear rate&$\gamma^\prime$&$2-6-20$\%\\
Contact stiffness number&$\kappa$&$400$\\
Inertial number&$I$&$ 10^{-6}-10^{-5}$\\
\hline
\multicolumn{3}{|c|}{Mechanical variables}\\
\hline
Density&$\rho$&$2000000\,kg/m^3$\\
Youngs modulus&$Y$&$70\,GPa$\\
Poissons ratio&$\nu$&$0,3$\\
Restitution coefficient&$e$&1\\
Rolling friction coefficient&$\mu_r$&$0-0,25-0,5-0,75-1$\\
Rolling viscous damping coefficient&$\eta_r$&$0-0,25-0,5-0,75$\\
Friction coefficient&$\mu_{p}$&$0,5$\\
\hline
\end{tabular}
\caption{DEM parameters used during simulations.}
\label{Influence rolling}
\end{table}

Then, the influence of the vertical pressure is investigated. The same set-up is used except the vertical pressure ($P=1\,MPa$ and $=100\,MPa$). The rolling friction coefficient is constant $\mu_r=0,5$ and the rolling viscous damping coefficient changes $\eta_r=0,25$, $0,5$ or $0,75$.

The different parameters needed for the DEM simulation are presented in table \ref{Influence rolling} and the value has been chosen from previous articles to represent rock material \cite{VandenEnde2018,Idrissi2020,Tang2016,Ferdowsi2020}.
We can notice the time step $dt$ must verify the Rayleigh condition \cite{Johnson1985,THORNTON1988133,Li2005} defined as:
\begin{equation}
dt_R = \pi\times r\times \frac{\sqrt{\rho / G}}{0,1631\times \nu + 0,8766}
\label{Time Step Check}
\end{equation}

With every computing test, the main problem is the running time. The time step $dt$ must be selected considering the number of particles, the computing power, the stability of the simulation and the time scale of the test. In our case, we are looking for a $10^2$ seconds term.  If we include the default value into equation \ref{Time Step Check} the time step is around $10^{-8}$ second and the running time skyrockets. To answer this we can easily change the density $\rho$ and the shear modulus $G$. We will see those parameters are included in two dimensionless numbers defined at the equation \ref{Dimensionless Numbers}: the contact stiffness number $\kappa$ \cite{Roux2002,DaCruz2005,Roux2010} and the inertial number $I$ \cite{Midi2004,DaCruz2005}.

\begin{equation}
\begin{array}{llr}
\kappa&=\left(\frac{Y}{P(1-\nu^2)}\right)^{2/3}&\text{The contact stiffness number}\\
I&=\gamma^\prime d_{50} \sqrt{\rho / P}&\text{The inertial number}
\end{array}
\label{Dimensionless Numbers}
\end{equation}

Where $\gamma^\prime=v_{shear}/h$ is the shear rate, $h$ is the height of the sample during the shear, $d_{50}$ the mean diameter, $P$ the pressure applied.

It appears the constitutive law is sensitive to $\kappa$ because grains are not rigid enough ($\kappa\leq 10^4$) \cite{Roux2002}. By this fact, it becomes not possible to change the Young modulus $Y$ (and so the shear modulus $G$). The inertial number $I$ represents the behavior of the grains, which can be associated with solids, liquids or gases \cite{Jaeger1996}. This dimensionless parameter does not affect the constitutive law if the flow regime is at critical state ($I\leq 10^{-3}$) \cite{Midi2004, DaCruz2005}. In conclusion, the density $\rho$ can be modified, if we stay under the condition $I\leq 10^{-3}$, to increase the time step and solve our computing problem.

\section{Results and discussion}

\emph{Rolling Parameters Influence}

\begin{figure}[h]
    \centering
    \includegraphics[width=0.9\linewidth]{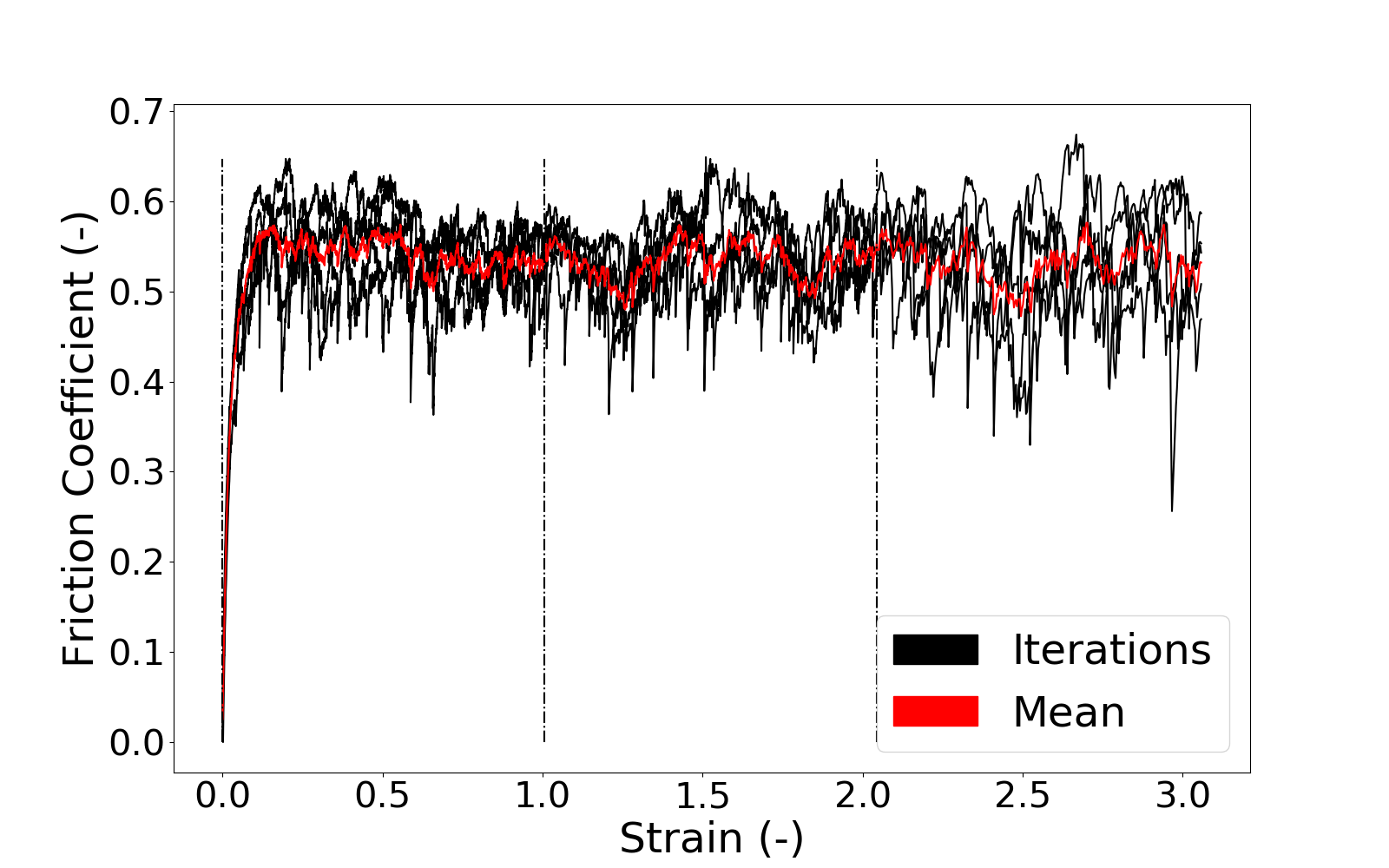}
    \caption{Example of $\mu$-$\gamma$ curve (dotted lines mark the different velocity steps 100 - 300 - 1000 $\mu$m/s).}
    \label{Mu Gamma}
\end{figure}

A parametric study has been done on the rolling friction coefficient $\mu_r$ and the rolling viscous damping coefficient $\eta_r$. As figure \ref{Mu Gamma} shows, the macro friction coefficient is plotted following the shear strain. This coefficient $\mu$ is computed by considering $\mu=F_y/F_x$, where $F_y$ (resp. $F_x$) is the component following the y-axis (resp. x-axis) of the force applied on the top plate. Because of the granular aspect, there is a lot of oscillation. To reduce this noise, at least 3 simulations are run by a set of parameters ($\mu_r$, $\eta_r$) and a mean curve is computed. Moreover, only the steady-state is considered and an average value is estimated. 

\begin{figure}[h!]
\centering
\includegraphics[width=0.9\linewidth]{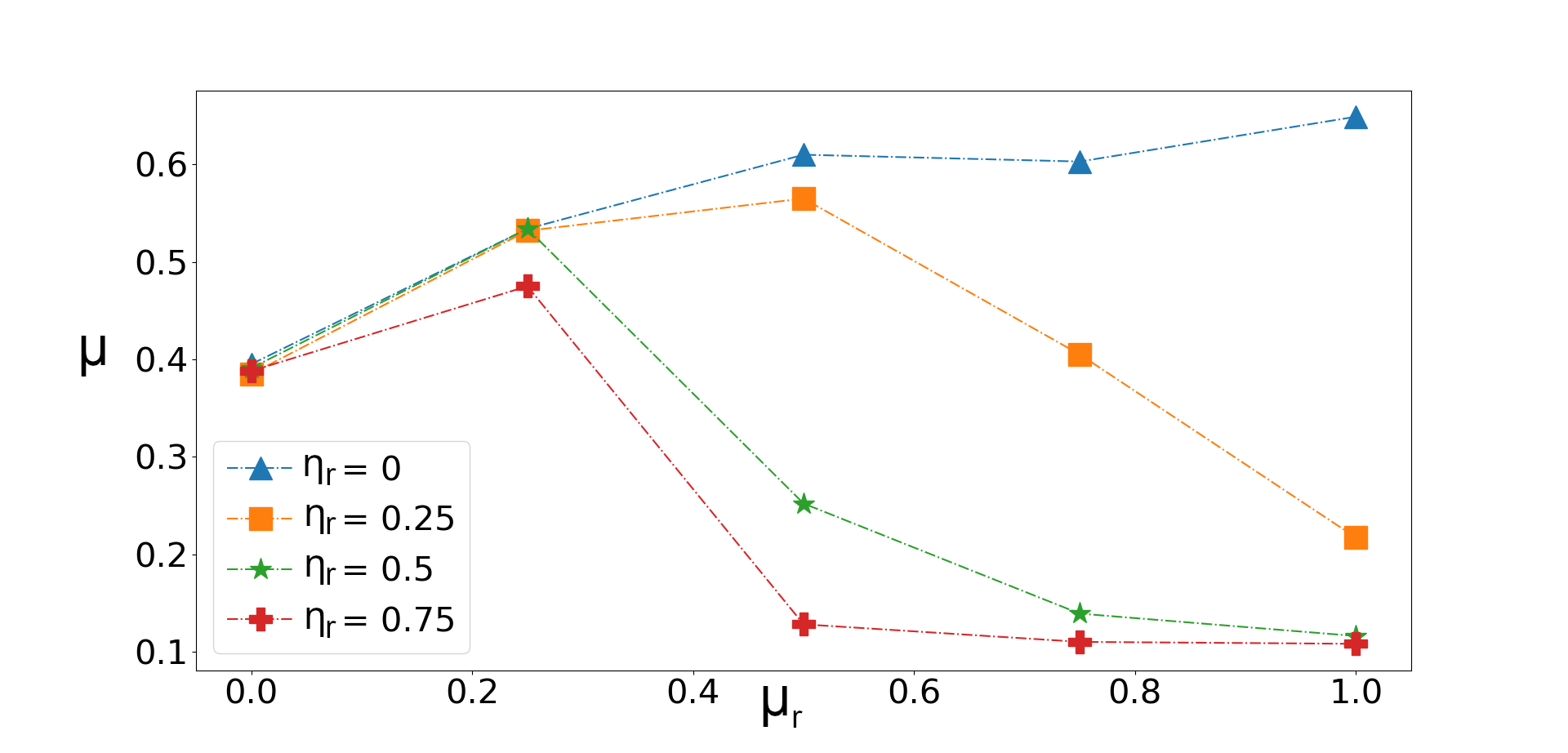}\\
$v_{shear}=100\,\mu m/s$ 

\includegraphics[width=0.9\linewidth]{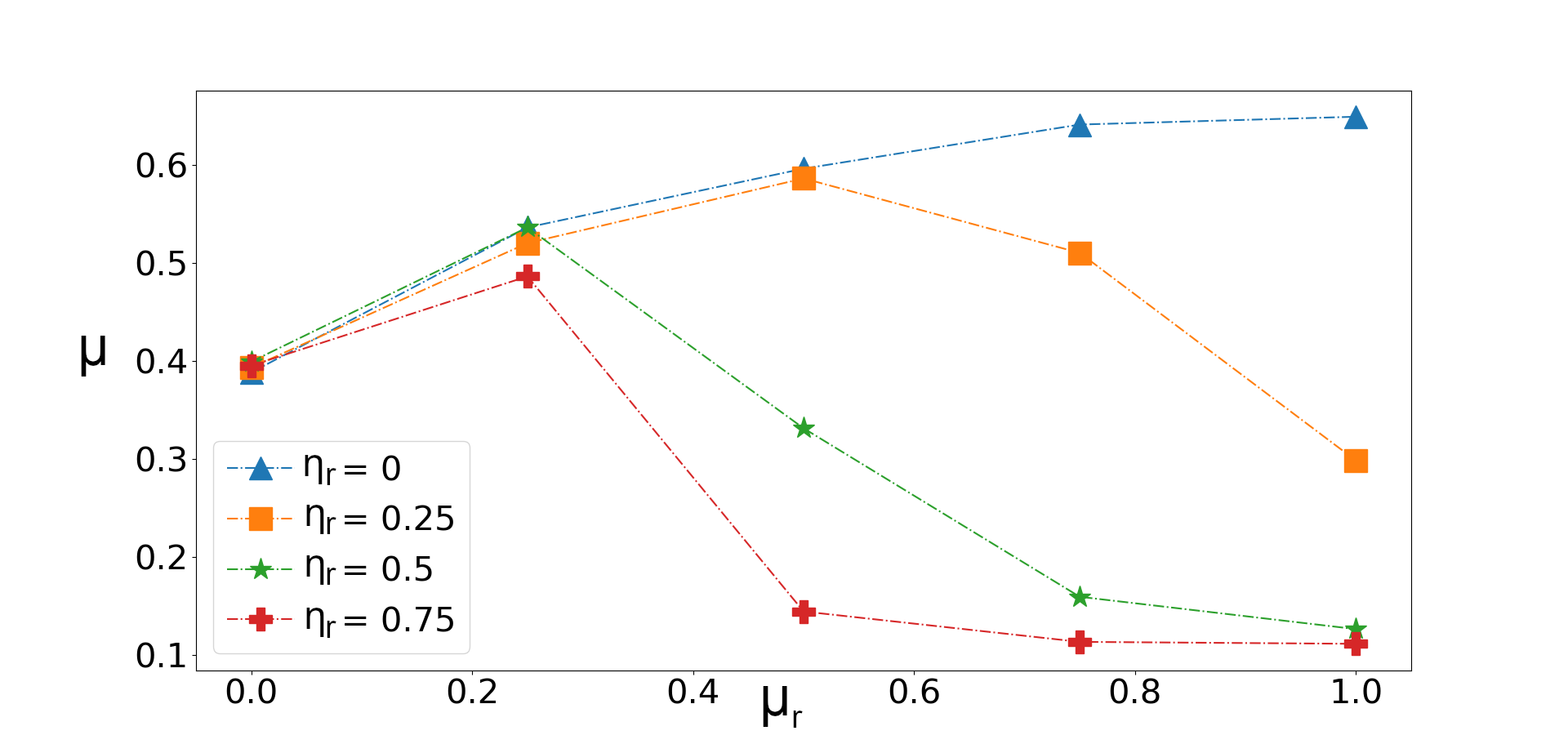}\\
$v_{shear}=300\,\mu m/s$

\includegraphics[width=0.9\linewidth]{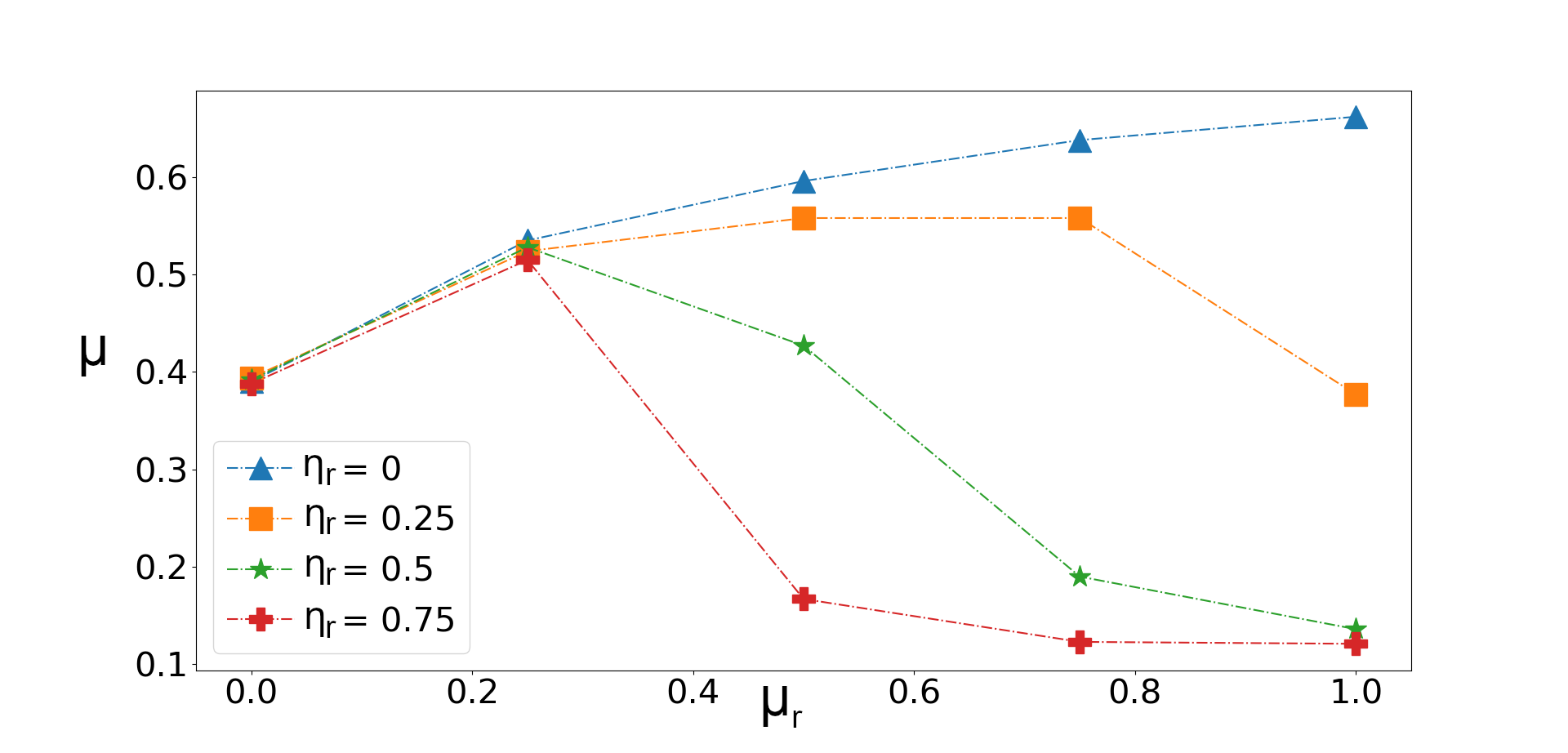}\\
$v_{shear}=1000\,\mu m/s$

\caption{Evolution of the macro friction coefficient with the rolling friction coefficient $\mu_r$ and the rolling viscous damping coefficient $\eta_r$ at different shear speeds with $P = 10 MPa$.}
\label{Gauge Friction Rolling Influence}
\end{figure}

The comparison of the macro friction coefficient with different parameters set is highlighted at figure \ref{Gauge Friction Rolling Influence}.  It appears there is an increase of the sheared layer friction coefficient with the rolling resistance $\mu_r$ until a critical point depending on the rolling damping $\eta_r$. This reduction of the stiffness with the rolling damping is not easy to understand at the first point. The larger is the damping, the stiffer should be the system. Two main questions should be answered: why does the friction coefficient decrease with the rolling resistance if there is damping? Why is the reduction larger with the damping value?

\vskip\baselineskip

\begin{figure}[h!]
    \centering
    \includegraphics[width=0.48\linewidth]{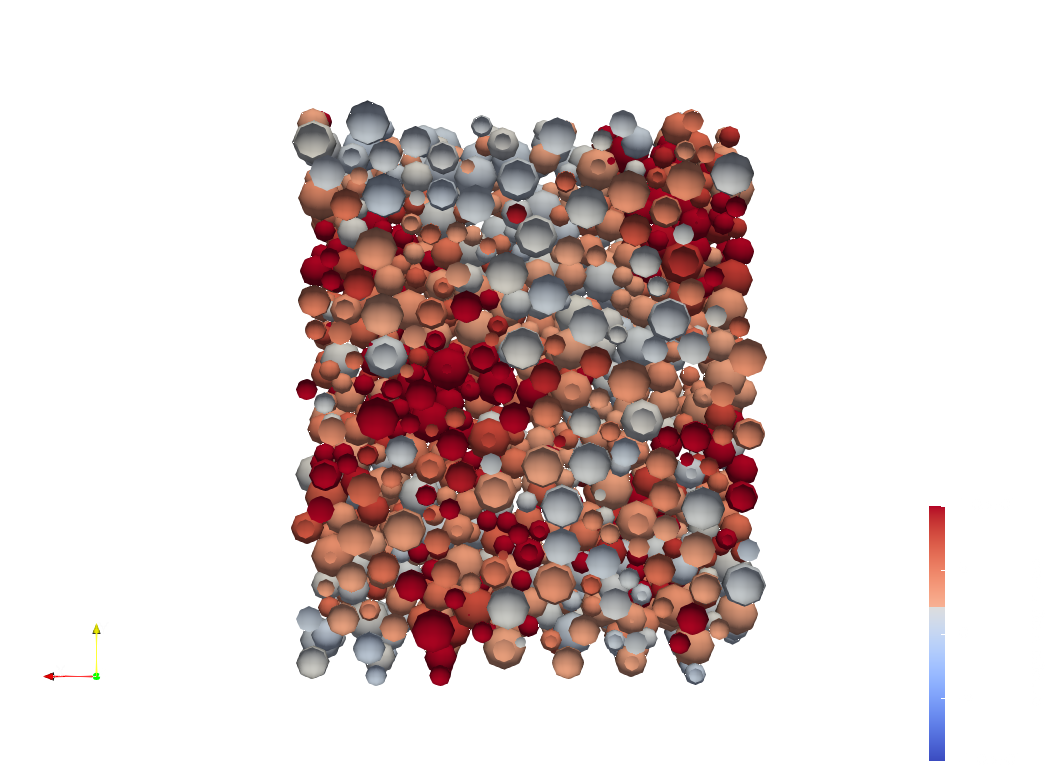} 
    \includegraphics[width=0.48\linewidth]{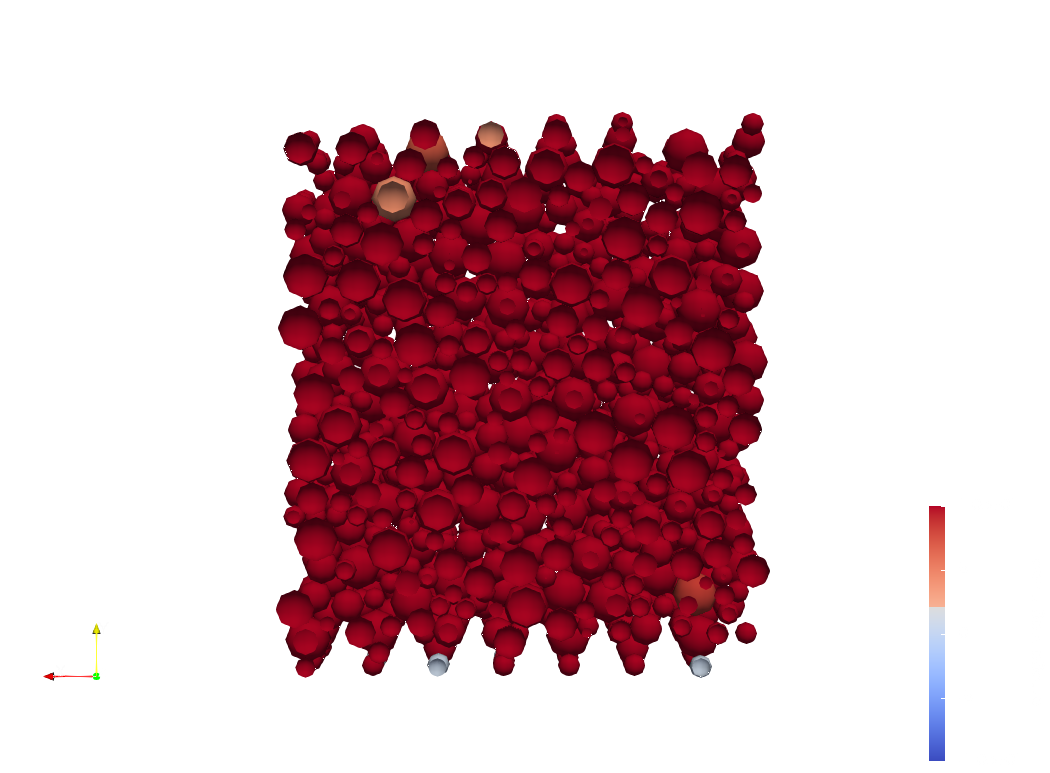}\\
    {\footnotesize $\mu_r=0,25-\eta_r=0,5$ \hspace{0.1\linewidth} $\mu_r=0,5-\eta_r=0,5$}
    
    \includegraphics[width=0.48\linewidth]{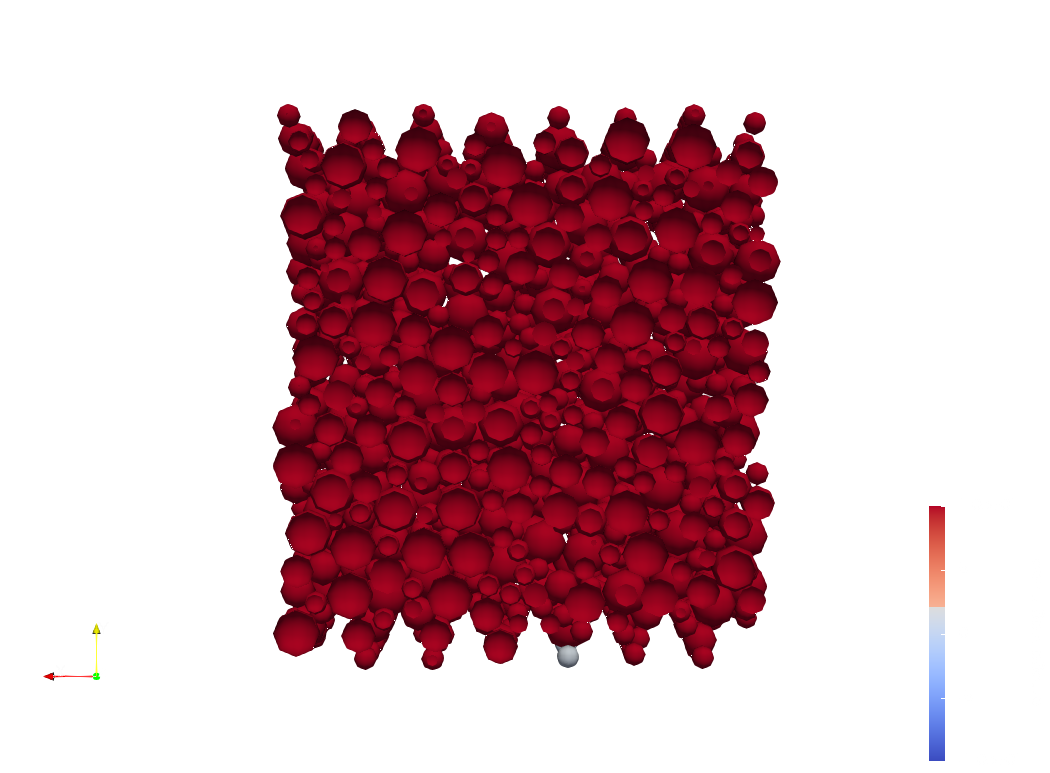}
    \includegraphics[width=0.48\linewidth]{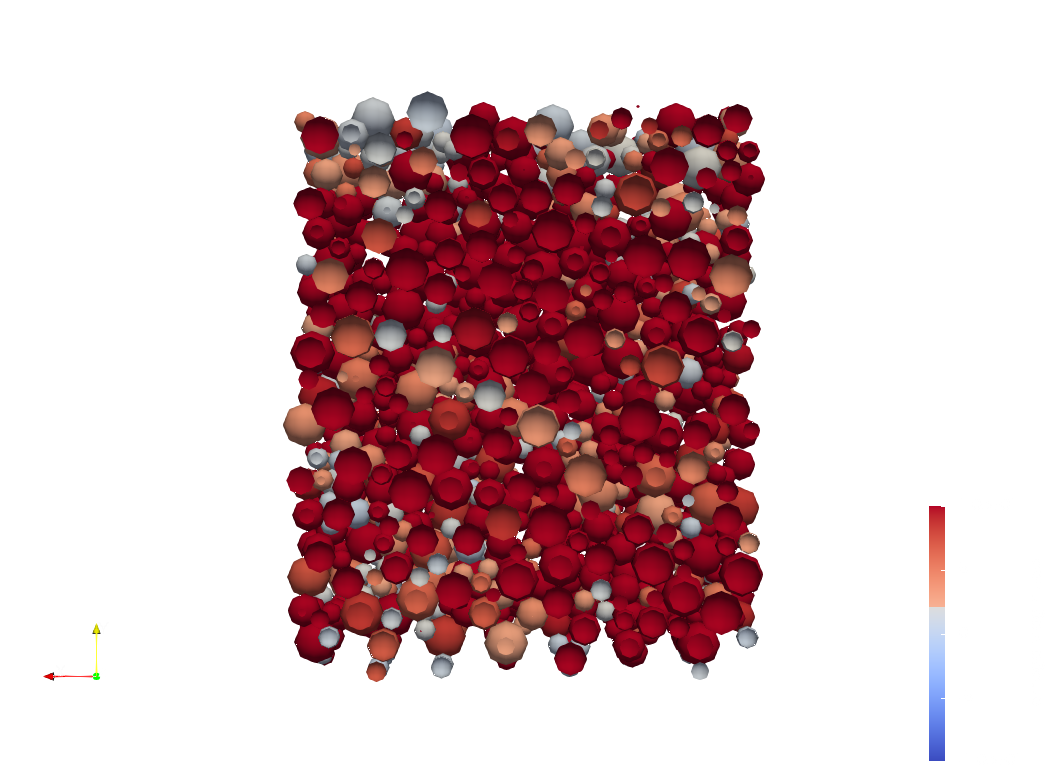}\\
    {\footnotesize $\mu_r=1-\eta_r=0,5$ \hspace{0.15\linewidth} $\mu_r=0,5-\eta_r=0$}
    
    \caption{Slide of sample highlighting grain rotation (rolling is in red) for several cases.}
    \label{Slide Rotation}
\end{figure}

Figure \ref{Slide Rotation} helps to understand the behavior. It shows the rotation of particle (in red) during four different cases. We can notice that the fewer rotations there are, the stiffer the system will be. It appears the number of rolling particle increases with the rolling resistance, see the three first plots of figure \ref{Slide Rotation}. The decrease of the friction coefficient is explained by particles rolling.
Moreover, it is shown at the second and last plot of figure \ref{Slide Rotation} that damping increases the number of rolling particles and so the friction coefficient is reduced. 

\vskip\baselineskip

A focus on the model equations must be done at relation \ref{Rolling Model Focus} to understand better those observations (the input rolling parameters are emphasized in red). First, it appears the increment of the spring $\Delta M_r^k$ depends on $\mu_r^2$ while the plastic limit $\mu_r R^*f_n$ depends only on $\mu_r$. There is a square factor between those values. Thus, this plastic limit, and so grain rolling, is reached easier with a larger rolling resistance $\mu_r$ for a same angular displacement $\theta$.

\begin{equation}
\begin{array}{ll}
M^m &= \text{\textcolor{red}{$\mu_r$}} R^*f_n \\ 
\Delta M_r^k &= -2,25k_n\text{\textcolor{red}{$\mu_r$}}^2R^{*2} \Delta \theta\\
M^d_{r,t+\Delta t} &=\left\lbrace
\begin{array}{r} 
\text{ if } \lvert M^k_{r,t+\Delta t}\rvert<\text{\textcolor{red}{$\mu_r$}} R^*f_n:\\
-2\text{\textcolor{red}{$\eta_r\mu_r$}} \sqrt{2,25I_rk_n}\omega \\
\\
\text{ if } \lvert M^k_{r,t+\Delta t}\rvert= \text{\textcolor{red}{$\mu_r$}} R^*f_n:\\
0
\end{array}\right.\\
\end{array}
\label{Rolling Model Focus}
\end{equation}

Concerning the damping, it avoids the variation of the angular position ($\Delta\omega \to 0$) during the elastic phase. As we have seen before, the main part of the sample is at the plastic phase and particles roll. So, it is as the damping acts in opposition of the angular spring, keeping grain into the plastic phase. We can notice that we have decided in this paper to shut down the damping moment when the angular plastic limit is reached (see equation \ref{Rolling Model Focus}).

In this way, we can understand better the reduction of the friction coefficient with the rolling stiffness $\mu_r$ if damping is active. We can notice there is no decrease but an increase of the friction coefficient in the case of no damping. In absence of this one, the angular spring can act normally. The larger the rolling parameter is, the stiffer the global sample is.

This observation can be used if we decide to track particle size distribution and grain shape evolution. Experiments and simulations have highlighted that particles tend to become less or more \cite{Zhu2021, Ueda2013, Zhang2018} rounded under large deformation.
The work of Buscarnera and Einav, extending the Continuum Breakage Mechanics, reconciles those conflicting observations \cite{Buscarnera2021}. Shape descriptors are converging to attractors. The evolution of the aspect ratio $\alpha$, related to the grain morphology, is plotted following a breakage parameter $B$ ($B=0$ means unbroken state and $B=1$ complete breakage) and the stress $\sigma$. It appears that from different initial values the aspect ratio converges to the the same value, the attractor. 

Softening and hardening behavior can be understood thank to relations between shape and rolling friction coefficient, the evolution of the grain shape and our work. For example, if the aspect ratio decreases during the test, particles become less rounded, the rolling friction coefficient increases, the sample toughness evolves depending on the position of the critical point (softening or hardening).

\vskip\baselineskip

\emph{Particle Size Distribution Influence}

As equation \ref{Rolling Model Focus} shows, the plastic moment depends on the equivalent radius $R^*$. Appendix \ref{Distribution Rolling Radius} notices smallest particles tend to roll more than the others. The particle size distribution should influence the macro friction coefficient.
Results from this observation are described at appendix \ref{Influence PSD}. This one highlights the fact that macro friction coefficient evolves with mean diameter $d_{50}$. There is an increase of the residual value with this parameter. Unfortunately, other simulations must be done to interpolate an accurate tendency. In facts, the number of particles (and so the height of the original sample) must be increased. Thus, particle size distribution with larger mean diameter $d_{50}$ can provide a sufficient number of particle to reduce the DEM noise. 

\vskip\baselineskip
\emph{Vertical Pressure Influence}

\begin{figure}[h!]
\centering

\includegraphics[width=0.9\linewidth]{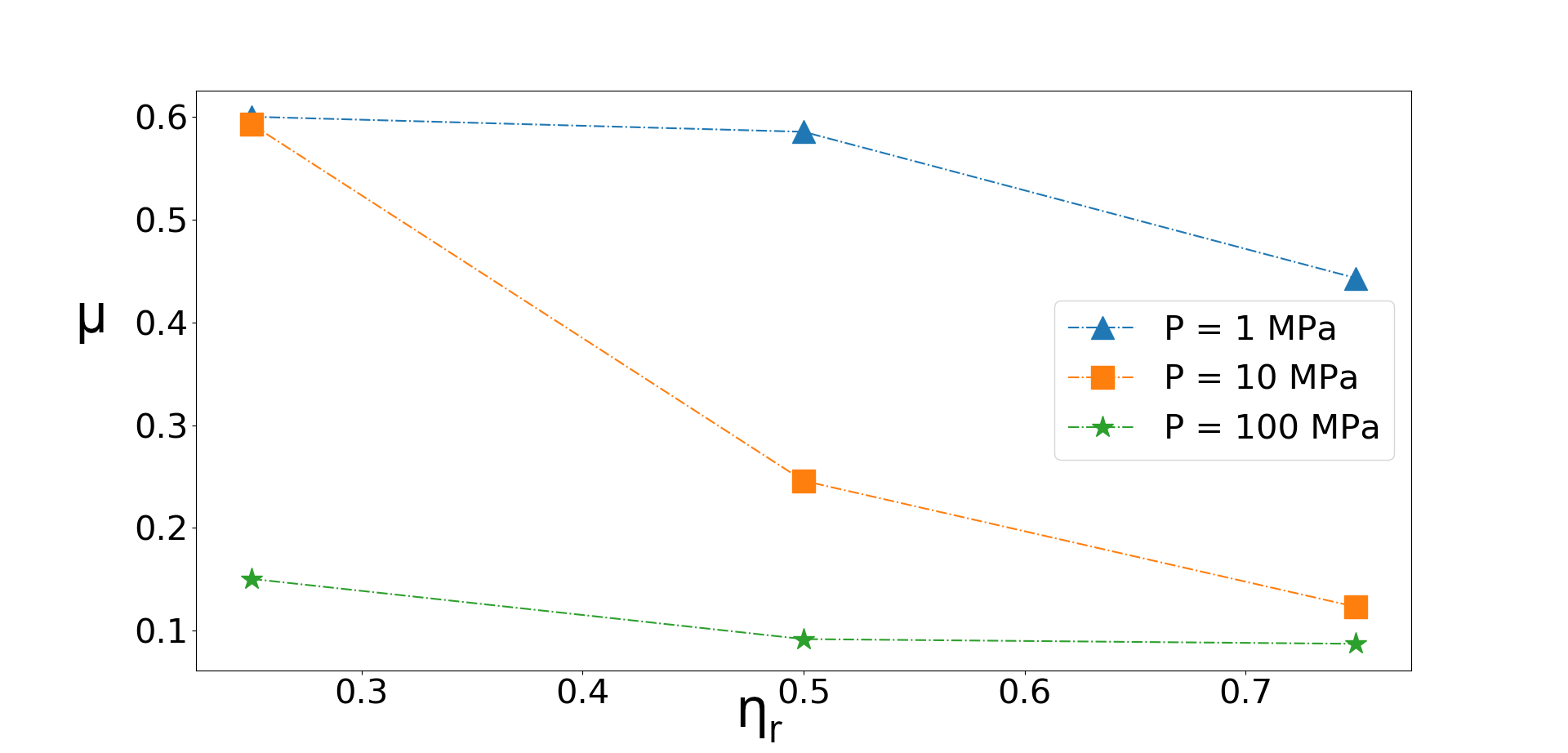}\\
$v_{shear}=100\,\mu m/s$

\includegraphics[width=0.9\linewidth]{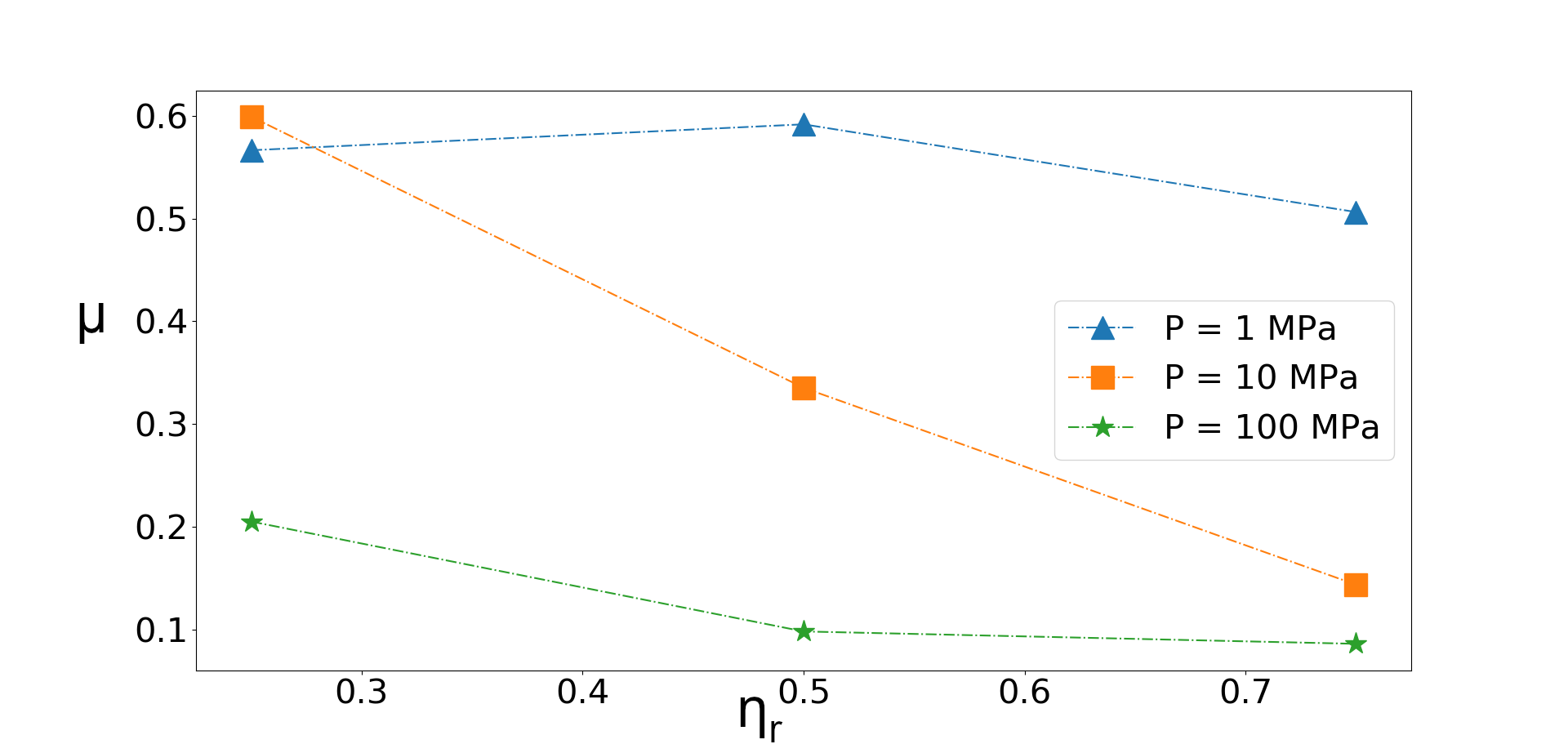}\\
$v_{shear}=300\,\mu m/s$

\includegraphics[width=0.9\linewidth]{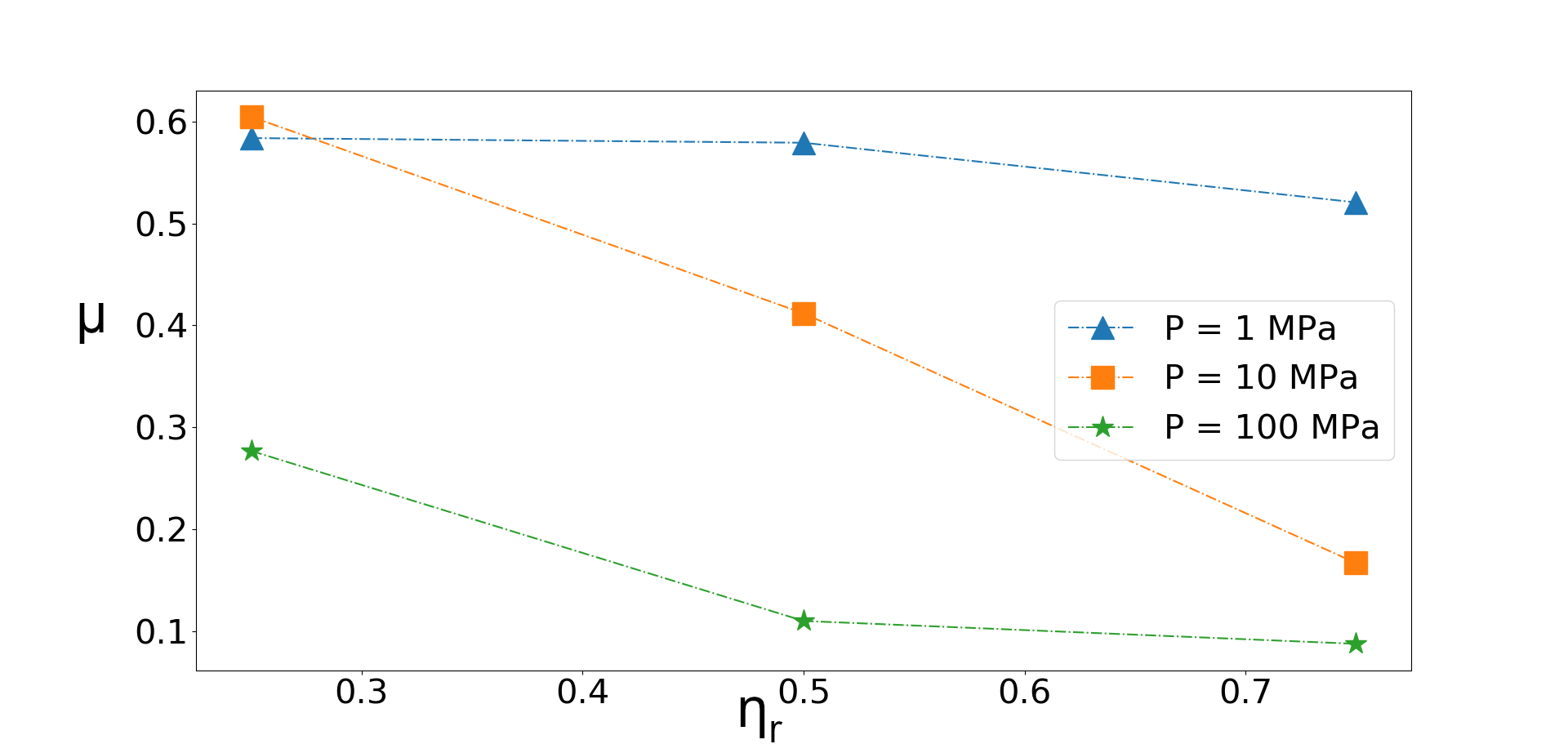}\\
$v_{shear}=1000\,\mu m/s$

\caption{Evolution of the macro friction coefficient with the rolling viscous damping coefficient $\eta_r$ and the vertical pressure $P$ at different shear speeds with $\mu_r=0,5$.}
\label{Gauge Friction Rolling Pressure Influence}
\end{figure}

Moreover, equation \ref{Rolling Model Focus} highlights the plastic moment depends on normal forces $f_n$ (and so on vertical pressure $P$). Figure \ref{Gauge Friction Rolling Pressure Influence} illustrates the evolution of the macro friction coefficient with different vertical pressure $P$ and rolling viscous damping coefficient $\eta_r$.

It appears the critical point defined previously (point before which the friction weakening appears) depends on the vertical pressure. This behaviour has already been observed \cite{Ferdowsi2020, Morgan1999}. It appears the importance of the rolling increases with the vertical pressure. The mean angular velocity $\omega_z$ on the z-axis have been computed over configurations $P-\eta_r-v_{shear}$ at table \ref{Mean Angular Velocity}. It is highlighted that the vertical pressure $P$ favours the grain rolling.
That is why a friction weakening occurs with this parameter.

\begin{table}[h]
    \centering
    
    \begin{tabular}{|r|c|c|c|}
    \hline
    \multicolumn{4}{|c|}{$v_{shear}=100\,\mu m/s$}\\
    \hline
    $\eta_r=$&$0,25$&$0,50$&$0,75$\\
    \hline
    $1\,MPa$&$0,3$&$2,2$&$33,7$\\
    $10\,MPa$&$20,3$&$455,6$&$1056,2$\\
    $100\,MPa$&$2384,5$&$7467,8$&$11774,1$\\
    \hline
    \end{tabular}
    
    \vskip\baselineskip
    
    \begin{tabular}{|r|c|c|c|}
    \hline
    \multicolumn{4}{|c|}{$v_{shear}=200\,\mu m/s$}\\
    \hline
    $\eta_r=$&$0,25$&$0,50$&$0,75$\\
    \hline
    $1\,MPa$&$0,6$&$3,6$&$34,1$\\
    $10\,MPa$&$23,9$&$426,6$&$1034,5$\\
    $100\,MPa$&$2327,3$&$7564,8$&$11685,2$\\
    \hline
    \end{tabular}  
    
    \vskip\baselineskip
    
    \begin{tabular}{|r|c|c|c|}
    \hline
    \multicolumn{4}{|c|}{$v_{shear}=1000\,\mu m/s$}\\
    \hline
    $\eta_r=$&$0,25$&$0,50$&$0,75$\\
    \hline
    $1\,MPa$&$1,3$&$4,9$&$29,7$\\
    $10\,MPa$&$28,1$&$342,1$&$1023,3$\\
    $100\,MPa$&$2070,0$&$7483,6$&$11759,4$\\
    \hline
    \end{tabular}    
    
    \vskip\baselineskip

    \caption{The mean angular velocity $\omega_z$ on the z-axis for different configurations $\eta_r-P$ at different shear speeds.}
    \label{Mean Angular Velocity}
\end{table}

\vskip\baselineskip
\emph{Speed Influence}

\begin{figure}[h!]
\centering
\includegraphics[width=0.8\linewidth]{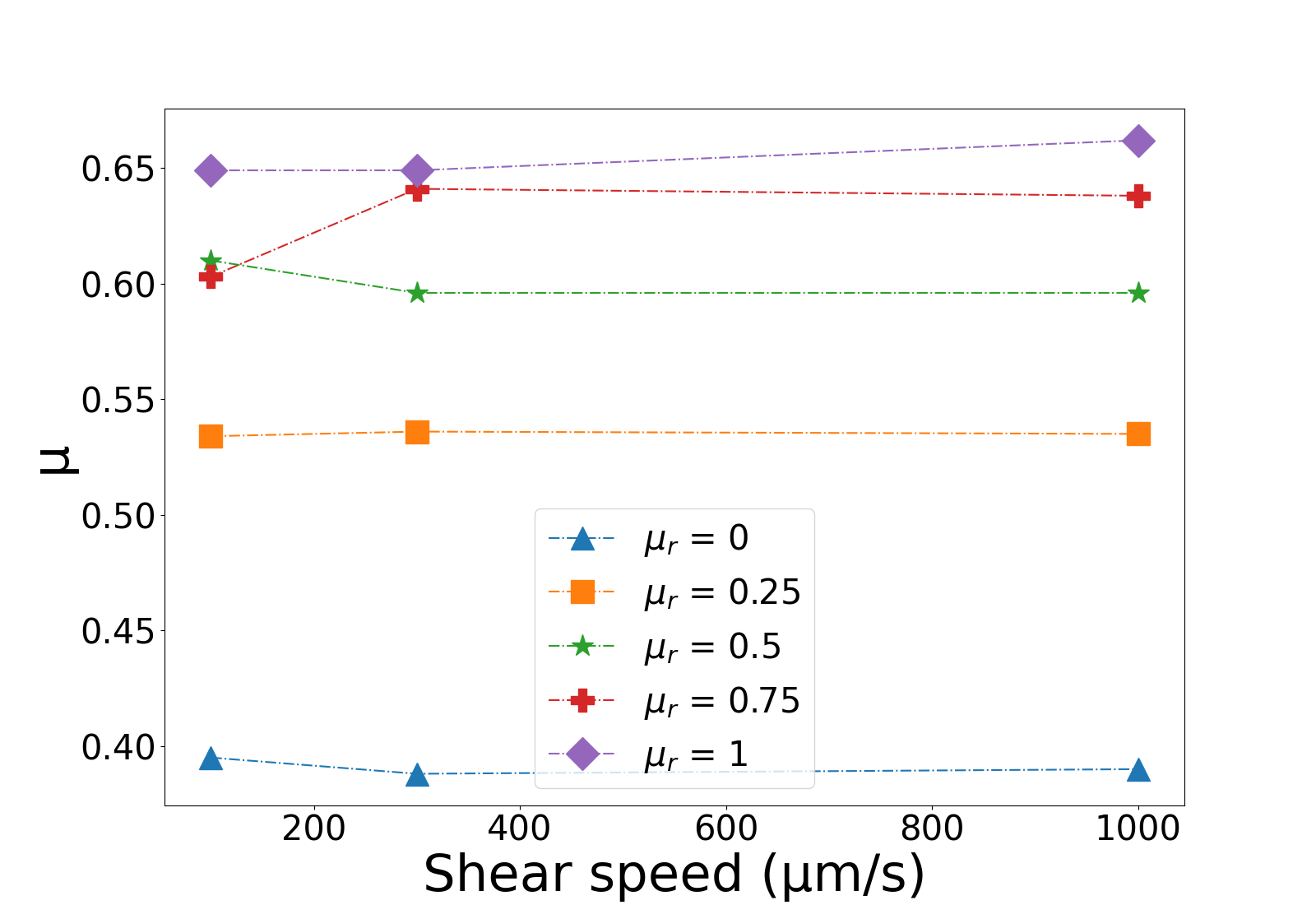}\\
$\eta_r=0$\\

\includegraphics[width=0.8\linewidth]{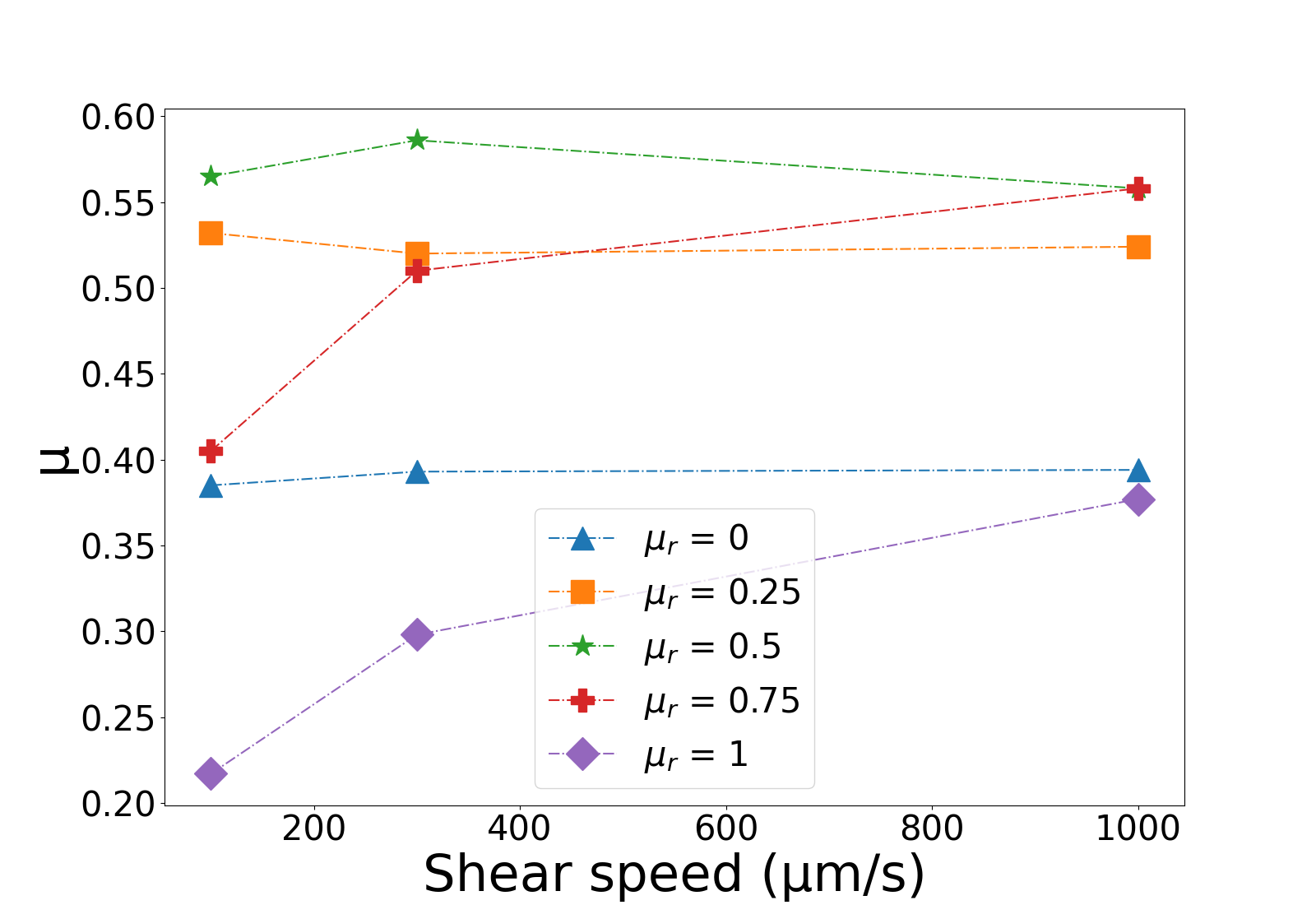}\\
$\eta_r=0,25$

\includegraphics[width=0.8\linewidth]{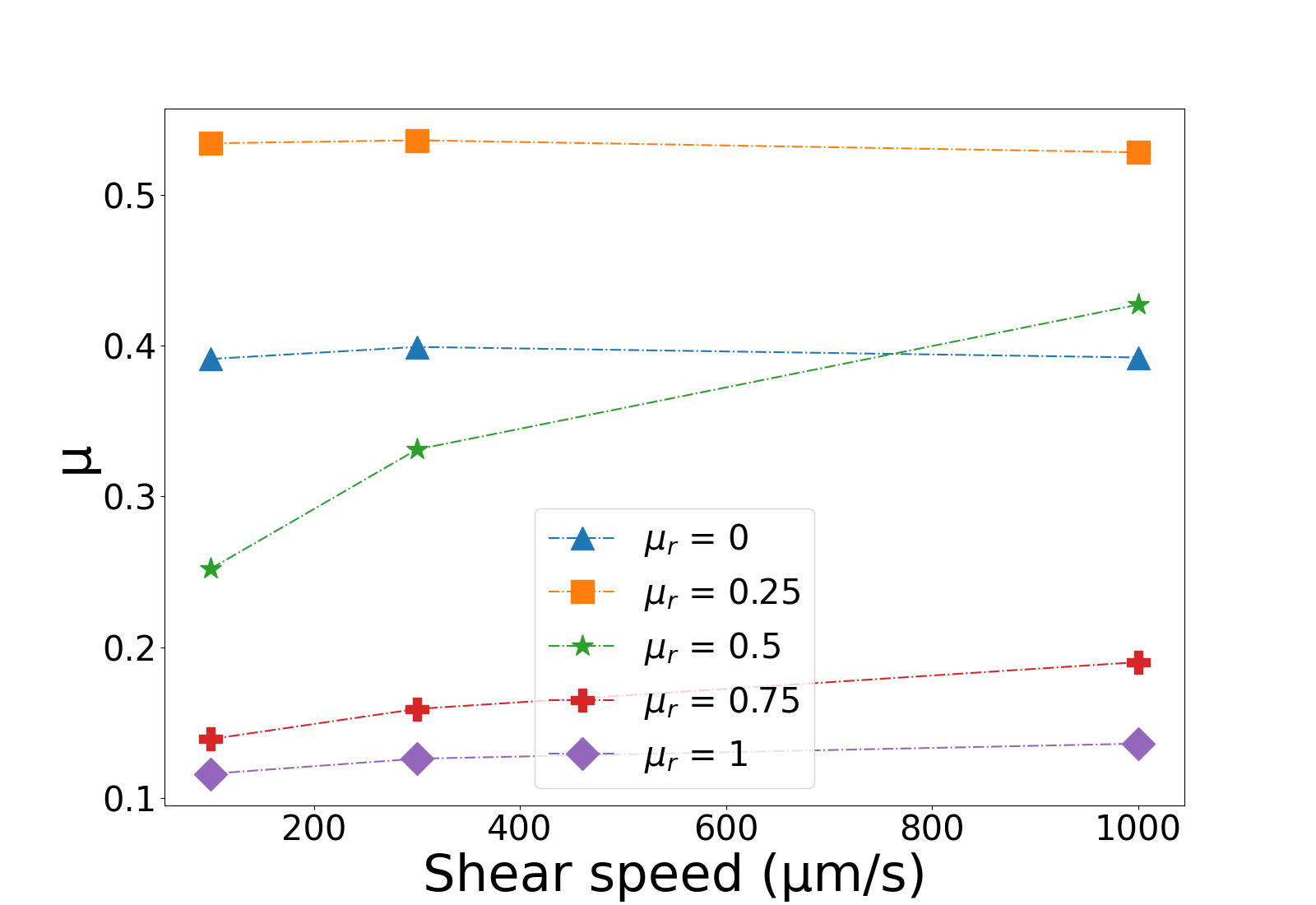}\\
$\eta_r=0,5$

\includegraphics[width=0.8\linewidth]{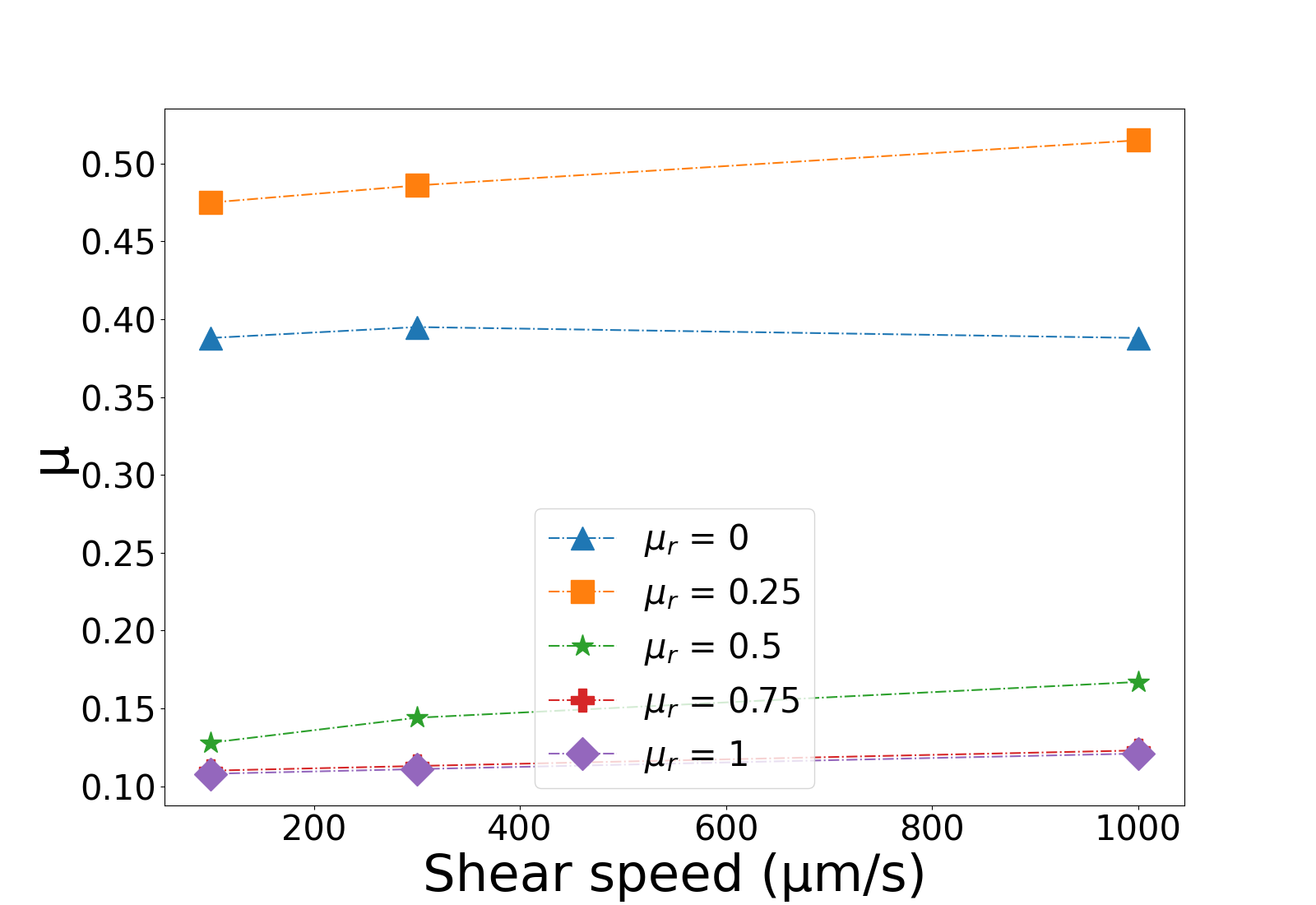}\\
$\eta_r=0,75$
\caption{Evolution of the sample friction coefficient with the shear speed and the rolling friction coefficient $\mu_r$ at different rolling viscous damping coefficients with $P=10\,MPa$.}
\label{Gauge Friction Rolling Influence Speed}
\end{figure}

Figure \ref{Gauge Friction Rolling Influence Speed} highlights the shearing speed influence on the system. It is the same results as before but plotted in another way. It appears there is no speed effect visible in most simulations as the friction coefficient keeps the same value. 
It is not surprising that no speed effects are spotted because there are no other parameters except the damping parameter which depends on speed or time.
A speed influence is nevertheless noticed for cases where the friction coefficient starts to decrease with rolling resistance (for example the case $\mu_r= 0,5$ and $\eta_r=0,5$ at figure \ref{Gauge Friction Rolling Influence Speed}). As shown at figure \ref{Slide Rotation} for this set, few particles (in orange or in white) are still not rolling during this critical step. The damping value is not large enough to cancel the effect of the spring and few grains are in the elastic phase. The damping creates so in this case a speed influence. If the damping value is larger, we have seen particles tend to be all in the plastic phase. If it is lower, the damping is negligible or null. In both cases, the speed effect disappears.

\section{Conclusion}

In this paper, we have considered granular materials into a plane shear configuration to investigate the effect of the rolling resistance and damping on the macroscopic friction coefficient. Thank numerical DEM simulations, the relation between those parameters becomes clearer.
It appears : 
\begin{enumerate}
    \item In the no damping case, the sample stiffness increases with the rolling resistance.
    \item The consideration of the rolling damping introduces a critical point. For a constant damping value, the sample stiffness increases the rolling parameter until this critical point is reached. Then, the stiffness starts to decrease until a residual value. Hence, the damping tend to act against the spring and grains roll.
    \item No visible speed effects have been highlighted except at critical point. For the same rolling resistance value : (i) When the damping parameter is not large enough, the angular spring is the main element and no speed dependency is spotted, (ii) when the damping parameter is too large, all grains are in the plastic phase (roll) and the residual value is reached and (iii) when the damping parameter is at critical value, there is no main element in the rolling model, speed dependency occurs. 
\end{enumerate}

\section{Acknowledgements}

Computational resources have been provided by the Consortium des Équipements de Calcul Intensif (CÉCI), funded by the Fonds de la Recherche Scientifique de Belgique (F.R.S.-FNRS) under Grant No. 2.5020.11 and by the Walloon Region. Support by the
CMMI-2042325 project is also acknowledged.

\section{Statements and Declarations}

\textbf{Conflict of interest} The authors declare that they have no conflict of interest.

\begin{appendices}

\section{Distribution of the rolling with the radius}\label{Distribution Rolling Radius}

\begin{figure}[h]
    \centering
    \includegraphics[width=0.8\linewidth]{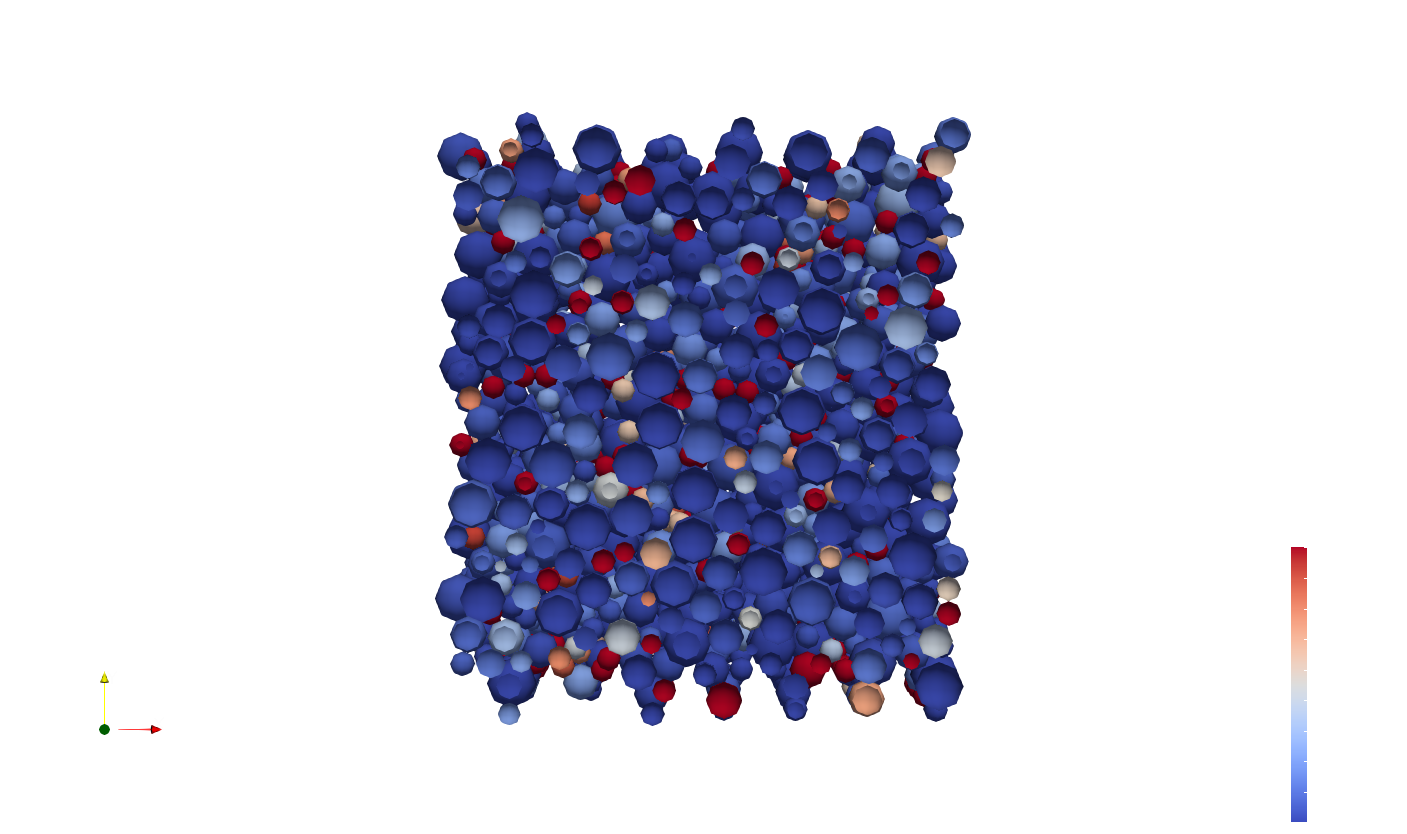}
    \caption{Distribution of the rolling in the sample (blue is small rolling, red is large rolling).}
    \label{Distribution Rolling}
\end{figure}

Figure \ref{Distribution Rolling} highlights the distribution of the rolling in the sample with $\mu_r=0,5$, $\eta_r=0,5$ and $v_{shear}=100\,\mu m/s$. It appears smallest particles are rolling more than largest particles. The observation stays qualitative and should be studied quantitatively to estimate better the behaviour.

\section{Influence of the Particle Size Distribution}\label{Influence PSD}

During the creation of a fault zone, the particles are crushed and the particle size distribution evolves greatly. Moreover, the figure \ref{Distribution Rolling} from the appendix \ref{Distribution Rolling Radius} highlights the fact that small particles tend to roll more than large ones (decreasing thus the sample toughness). To investigate how this evolution can affect the frictional behavior of a fault, the Particle Size Distribution (PSD) is described by two main parameters: the mean diameter $d_{50}$ and the fractal dimension $D$ defined at equation \ref{Fractal Number} \cite{Sammis1987, Einav2007, Rattez2021}.

\begin{equation}
    \label{Fractal Number}
    N = r^{-D}
\end{equation}
with $N$ is the number of particle of radius $r$.

\begin{figure}[h]
    \centering
    \includegraphics[width=0.9\linewidth]{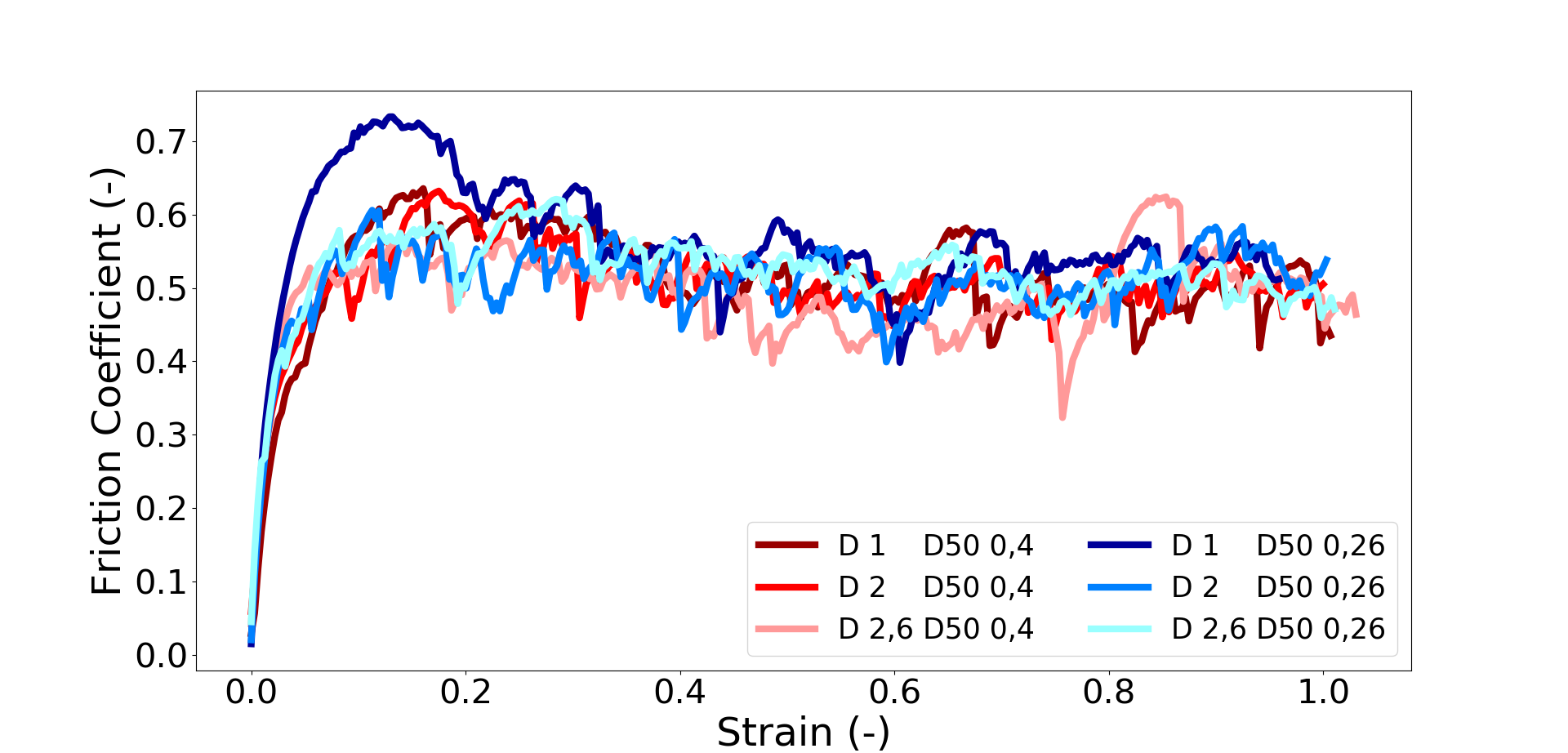}\\
    $\mu_r = 0.2$
    
    \includegraphics[width=0.9\linewidth]{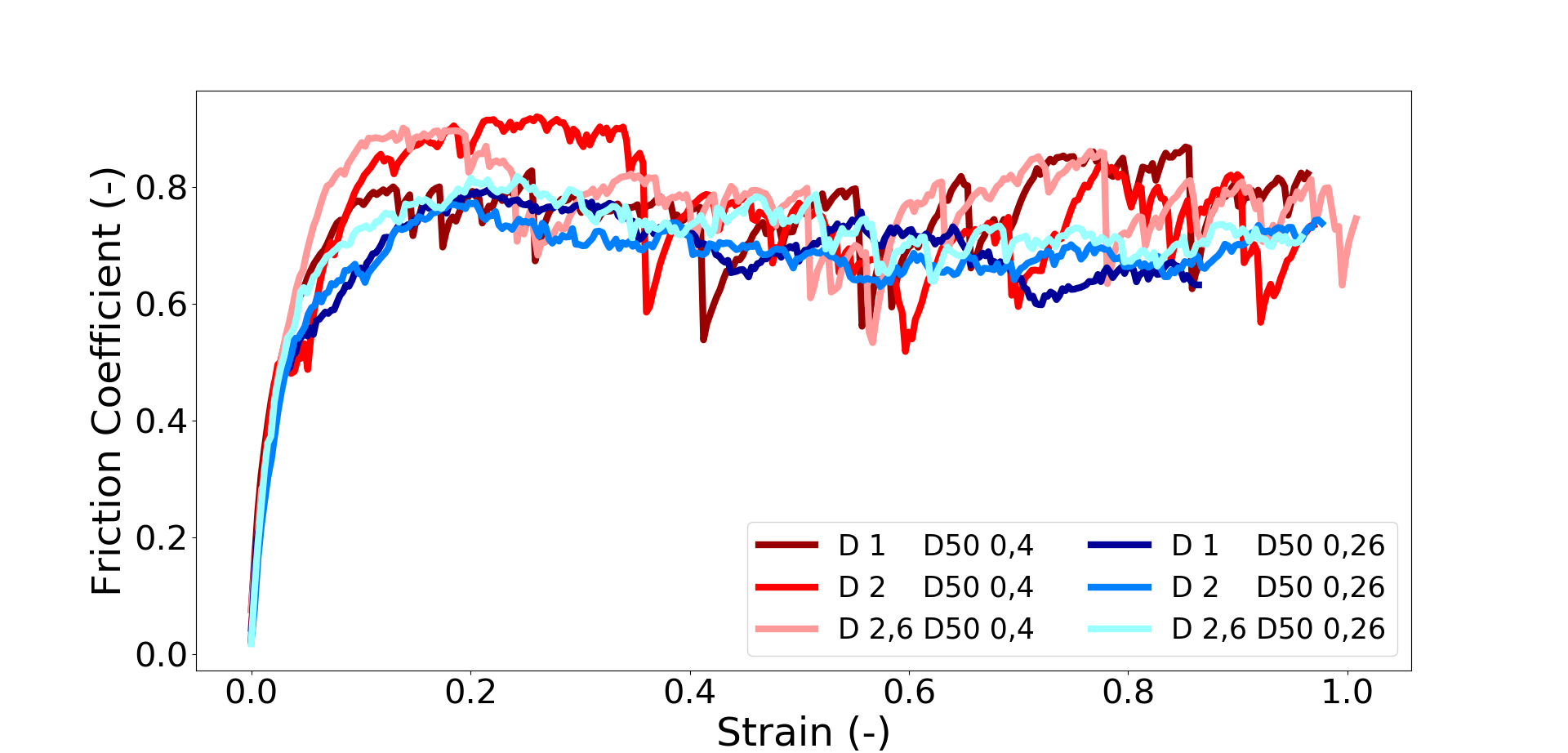}\\
    $\mu_r = 1.4$
    \caption{Evolution of the $\mu$ with the strain for several PSD.}
    \label{Influence mu PSD}
\end{figure}

Simulations have been run at the shear speed of $1000\,\mu m/s$ during one step. Two rolling resistance $\mu_r$ are investigated $0,2$ (smooth particles) and $1,4$ (rough particles) and the rolling damping $\eta_r$ is not considered ($=0$). The different PSD used are described in table \ref{PSD} and results are shown in figure \ref{Influence mu PSD}.

\begin{table}[h]
\centering
\begin{tabular}{|p{0.15\linewidth}|l|l|p{0.18\linewidth}|}
\hline
PSD Name&Radius&Percentage&Number of particles\\
\hline
D 1 & R1 $= 0,25\,mm$ & $27\%$ & 750 \\
D50 0,4 & R2 $= 0,21\,mm$ & $32\%$ & \\
& R3 $= 0,16\,mm$ & $41\%$ & \\
\hline
D 2 & R1 $= 0,25\,mm$ & $22\%$ & 750 \\
D50 0,4 & R2 $= 0,21\,mm$ & $31\%$ & \\ 
& R3 $= 0,17\,mm$ & $47\%$ & \\ 
\hline
D 2,6 & R1 $= 0,25\,mm$ & $20\%$ & 750 \\ 
D50 0,4 & R2 $= 0,21\,mm$ & $30\%$ & \\
& R3 $= 0,17\,mm$ & $50\%$ & \\
\hline
D 1 & R1 $= 0,2\,mm$ & $22\%$  & 2500 \\
D50 0,26 & R2 $= 0,14\,mm$ & $30\%$ & \\
& R3 $= 0,09\,mm$ & $48\%$ & \\
\hline
D 2 & R1 $= 0,2\,mm$ & $14\%$ & 2500 \\ 
D50 0,26 & R2 $= 0,15\,mm$ & $29\%$ & \\
& R3 $= 0,1\,mm$ & $57\%$ & \\
\hline
D 2,6 & R1 $= 0,2\,mm$ & $12\%$ & 2500 \\
D50 0,26 & R2 $= 0,15\,mm$ & $25\%$ & \\
& R3 $= 0,11\,mm$ & $63\%$ & \\
\hline
\end{tabular}
\caption{Distribution used described by discrete radius, percentage of the mass and total number of grains in the case of smooth and rough particles.}
\label{PSD}
\end{table}

In the case of smooth particle, it seems curves are really similar. The only exception is the case D 1 D50 0,26 where a peak appears. More differences can be appreciated in the case of rough particles. A peak can be shown at the start of the shear movement. This peak varies with the PSD, for example, the D 2,6 D50 0,4 one is more pronounced than the D 1 D50 0,4 one. It has already been observed that the peak tends to increase with the fractal dimension \cite{Rattez2020}.

PSD has a significant influence on the
residual friction coefficient if the rolling resistance becomes large. In fact, this parameter seems to stay around the value of 0,5 for smooth particles. Whereas, in the case of rough particles, values from tests with a mean diameter $d_{50} = 0,4\,mm$ are larger ($\approx 0.8$) than the ones obtained from tests with $d_{50} = 0,26\,mm$ ($\approx 0.7$). 

In conclusion, it appears the rolling resistance has an influence on the friction coefficient via the PSD (and especially via the mean diameter $d_{50}$). More tests must be done to appreciate results. Hence, curves obtained from PSD with $d_{50}=0.4\,mm$ are really noisy because there are not enough grains (the sample height  have been conserved).

\end{appendices}


\begin{thebibliography}{99}

\footnotesize

\bibitem{MYERS2004947} Myers, R., Aydin, A.: The evolution of faults formed by shearing across joint zones in sandstone. J. of Struct. Geol. 26, 947-966 (2004) https://doi.org/10.1016/j.jsg.2003.07.008

\bibitem{Poulet2014} Poulet, T., Veveakis, M., Herwegh, M., Buckingham, T., Regenauer-Lieb, K.: Modeling episodic fluid-release events in the ductile carbonates of the Glarus thrust. Geophys. Res. Lett. 41, 7121-7128 (2014). https://doi.org/10.1002/2014GL061715

\bibitem{Segui2020} Segui, C., Rattez, H., Veveakis, M.: On the stability of deep-seated landslides. The cases of Vaiont (Italy) and Shuping (Three Gorges Dam, China). J. of Geophys. Res.: Earth Surf. 125, e2019JF005203 (2020). https://doi.org/10.1029/2019JF005203

\bibitem{https://agupubs.onlinelibrary.wiley.com/doi/pdf/10.1029/97RG00426} Iverson, R.M.: The physics of debris flows. Rev. Geophys. 35, 245–296 (1997). https://doi.org/10.1029/97RG00426

\bibitem{SulemVardoulakis1995} Sulem, J., Vardoulakis, I.G.: Bifurcation Analysis in Geomechanics (1st ed.). CRC Press, London (1995). https://doi.org/10.1201/9781482269383

\bibitem{Sulemetal2011} Sulem, J., Stefanou, I., Veveakis, M.: Stability analysis of undrained adiabatic shearing of a rock layer with Cosserat microstructure. Granul. Matter 13, 261-268 (2011). https://doi.org/10.1007/s00603-018-1529-7

\bibitem{Rattez2018a} Rattez, H., Stefanou, I., Sulem, J.: The importance of Thermo-Hydro-Mechanical couplings and microstructure to strain localization in 3D continua with application to seismic faults. Part I: Theory and Linear Stability Analysis. J. of the Mech. and Phys. of Solids 115, 54-76 (2018). https://doi.org/10.1016/j.jmps.2018.03.004

\bibitem{Rattez2018b} Rattez, H., Stefanou, I., Sulem, J., Veveakis, M., Poulet, T.: The importance of Thermo-Hydro-Mechanical couplings and microstructure to strain localization in 3D continua with application to seismic faults. Part II: Numerical implementation and post-bifurcation analysis. J. of the Mech. and Phys. of Solids 115, 1-29 (2018). https://doi.org/10.1016/j.jmps.2018.03.003

\bibitem{Rattez2018c} Rattez, H., Stefanou, I., Sulem, J., Veveakis, M., Poulet, T.: Numerical Analysis of Strain Localization in Rocks with Thermo-hydro-mechanical Couplings Using Cosserat Continuum. Rock Mech. and Rock Eng. 51, 3295–3311 (2018). https://doi.org/10.1007/s00603-018-1529-7

\bibitem{https://agupubs.onlinelibrary.wiley.com/doi/10.1029/2022JB025209?af=R} Papachristos, E., Stefanou, I., Sulem, J.: A discrete elements study of the frictional behavior of fault gouges. J. of Geophys. Res.: Solid Earth 128, e2022JB025209 (2023). https://doi.org/10.1029/2022JB025209

\bibitem{Burman1980} Burman, B.C., Cundall, P.A., Strack, O.D.L.: A discrete numerical model for granular assemblies. Geotech. 30, 331-336 (1980). https://doi.org/10.1680/geot.1980.30.3.331

\bibitem{PhysRevE.89.042210} Imole, O.I., Wojtkowski, M., Magnanimo, V., Luding, S.: Micro-macro correlations and anisotropy in granular assemblies under uniaxial loading and unloading. Phys. Rev. E 89, 042210 (2014). https://doi.org/10.1103/PhysRevE.89.042210

\bibitem{PhysRevE.92.022202} Gonz\'alez, S., Windows-Yule, C.R.K., Luding, S., Parker, D.J., Thornton, A.R.: Forced axial segregation in axially inhomogeneous rotating systems. Phys. Rev. E 92, 022202 (2015). https://doi.org/10.1103/PhysRevE.92.022202

\bibitem{OSullivan2011} O'Sullivan, C.: Particulate Discrete Element Modelling: A Geomechanics Perspective. CRC Press, London (2011). https://doi.org/10.1201/9781482266498

\bibitem{HANLEY20151100} Hanley, K.J., O'Sullivan, C., Huang, X.: Particle-scale mechanics of sand crushing in compression and shearing using DEM. Soils and Found. 55, 1100-1112 (2015). https://doi.org/10.1016/j.sandf.2015.09.011

\bibitem{Zhang2021} Zhang, N., Ciantia, M.O., Arroyo, M., Gens, A.: A contact model for rough crushable sand. Soils and Found. 61, 798-814 (2021). https://doi.org/10.1016/j.sandf.2021.03.002

\bibitem{Rutter1976} Elliott, D., Rutter, E.: The Kinetics of Rock Deformation by Pressure Solution. Philos. Trans. of The R. Soc. A: Math., Phys. and Eng. Sci. 283, 218-219 (1976). https://doi.org/10.1098/rsta.1976.0079

\bibitem{LEHNER1995153} Florian, K.L.: A model for intergranular pressure solution in open systems. Tectonophys. 245, 153-170 (1995). https://doi.org/10.1016/0040-1951(94)00232-X

\bibitem{VandenEnde2018} van den Ende, M.P.A., Marketos, G., Niemeijer, A.R., Spiers, C.J.: Investigating Compaction by Intergranular Pressure Solution Using the Discrete Element Method. J. of Geophys. Res.: Solid Earth 123, 107-124 (2018). https://doi.org/10.1002/2017JB014440

\bibitem{Abe2002} Abe, S., Dieterich, J.H., Mora, P., Place, D.: Simulation of the influence of rate- and state-dependent friction on the macroscopic behavior of complex fault zones with the lattice solid model. Pure and Appl. Geophys. 159, 1967-1983 (2002). https://doi.org/10.1007/s00024-002-8718-7

\bibitem{Morgan2004} Morgan, J.K.: Particle dynamics simulations of rate- and state-dependent frictional sliding of granular fault gouge. Pure and Appl. Geophys. 161, 1877-1891 (2004). https://doi.org/10.1007/s00024-004-2537-y

\bibitem{Potyondy2004} Potyondy, D.O., Cundall, P.A.: A bonded-particle model for rock. Int. J. of Rock Mech. and Min. Sci. 41, 1329-1364 (2004). https://doi.org/10.1016/j.ijrmms.2004.09.011

\bibitem{Li2017} Zhao, H., Sang, Y., Deng, A., Ge, L.: Influences of Stiffness Ratio, Friction Coefficient and Strength Ratio on the Macro Behavior of Cemented Sand Based on DEM. In: Li, X., Feng, Y., Mustoe, G. (eds) DEM 2016: Proceedings of the 7th International Conference on Discrete Element Methods, pp. 485-495. Springer, Singapore (2017). https://doi.org/10.1007/978-981-10-1926-5\_51

\bibitem{Casas2020} Casas, N., Mollon, G., Daouadji, A.: Cohesion and Initial Porosity of Granular Fault Gouges control the Breakdown Energy and the Friction Law at the Onset of Sliding. ESS Open Archive (2020). https://doi.org/10.1002/essoar.10504966.1

\bibitem{Soulie2006} Soulié, F., El Youssoufi, M.S., Cherblanc, F., Saix, C.: Capillary cohesion and mechanical strength of polydisperse granular materials. The European Phys. J. E 21, 349–357 (2006). https://doi.org/10.1140/epje/i2006-10076-2

\bibitem{Dorostkar2018} Dorostkar, O., Guyer, R.A., Johnson, P.A.,  Marone, C., Carmeliet, J.: Cohesion-Induced Stabilization in Stick-Slip Dynamics of Weakly Wet, Sheared Granular Fault Gouge. J. of Geophys. Res.: Solid Earth 123, 2115-2126 (2018). https://doi.org/10.1002/2017JB015171

\bibitem{doi:10.1680/geot.2002.52.3.157} Vardoulakis, I.: Dynamic thermo-poro-mechanical analysis of catastrophic landslides. Géotech. 52, 157-171 (2002). https://doi.org/10.1680/geot.2002.52.3.157

\bibitem{Rice2006} Rice, J.R.: Heating and weakening of faults during earthquake slip. J. of Geophys.l Res.: Solid Earth 111, B05311 (2006). https://doi.org/10.1029/2005JB004006

\bibitem{Gan2012} Gan, Y., Rognon, P., Einav, I.: Phase transitions and cyclic pseudotachylyte formation in simulated faults. Philos. Mag. 92, 3405-3417 (2012). https://doi.org/10.1080/14786435.2012.669062

\bibitem{Mollon2021} Mollon, G., Aubry, J., Schubnel, A.: Simulating melting in 2D seismic fault gouge. J. of Geophys. Res.: Solid Earth 126, 6 (2021). https://doi.org/10.1029/2020JB021485

\bibitem{Idrissi2020} Idrissi, H., Samaee, V., Lumbeeck, G., van der Werf, T., Pardoen, T., Schryvers, D., Cordier, P.: In Situ Quantitative Tensile Testing of Antigorite in a Transmission Electron Microscope. J. of Geophys. Res.: Solid Earth 125, 1-12 (2020). https://doi.org/10.1029/2019JB018383

\bibitem{Midi2004} Midi GDR: On dense granular flows. European Phys. J. E 14, 341-365 (2004). https://doi.org/10.1140/epje/i2003-10153-0

\bibitem{doi:10.1680/geot.2004.54.8.539} Oda, M., Takemura, T., Takahashi, M.: Microstructure in shear band observed by microfocus X-ray computed tomography. Géotech. 54, 539-542 (2004). https://doi.org/10.1680/geot.2004.54.8.539

\bibitem{ODA1982269} Oda, M., Konishi, J., Nemat-Nasser, S.: Experimental micromechanical evaluation of strength of granular materials: Effects of particle rolling. Mech. of Mater. 1, 269-283 (1982). https://doi.org/10.1016/0167-6636(82)90027-8

\bibitem{Zhou1999} Zhou, Y.C., Wright, B.D., Yang, R.Y., Xu, B.H., Yu, A.B.: Rolling friction in the dynamic simulation of sandpile formation. Phys. A: Stat. Mech. and its Appl. 269, 536-553 (1999). https://doi.org/10.1016/S0378-4371(99)00183-1

\bibitem{Alonso-Marroquin2006} Alonso-Marroquin, F., Vardoulakis, I., Herrmann, H.J., Weatherley, D., Mora, P.: Effect of rolling on dissipation in fault gouges. Phys. Rev. E - Stat., Nonlinear, and Soft Matter Phys. 74, 1-10  (2006). https://doi.org/10.1103/PhysRevE.74.031306

\bibitem{Papanicolopulos2011} Papanicolopulos, S.A., Veveakis, E.: Sliding and rolling dissipation in Cosserat plasticity. Granul. Matter 13, 197-204 (2011). https://doi.org/10.1007/s10035-011-0253-8

\bibitem{Ai2011} Ai, J., Chen, J.F., Rotter, J.M., Ooi, J.Y.: Assessment of rolling resistance models in discrete element simulations. Powder Technol. 206, 269-282 (2011). https://doi.org/10.1016/j.powtec.2010.09.030

\bibitem{Zhao2016} Zhao, C., Li, C.: Influence of rolling resistance on the shear curve of granular particles. Phys. A: Stat. Mech. and its Appl. 460, 44-53 (2016). https://doi.org/10.1016/j.physa.2016.04.043

\bibitem{IwashitaK.Oda1998} Iwashita, K., Oda, M.: Rolling Resistance At Contacts in Simulation of Shear Band. Asce 124, 285-292 (1998). https://doi.org/10.1061/(ASCE)0733-9399(1998)124:3(285)

\bibitem{Iwashita2000} Iwashita, K., Oda, M.: Micro-deformation mechanism of shear banding process based on modified distinct element method. Powder Technol. 109, 192-205 (2000). https://doi.org/10.1016/S0032-5910(99)00236-3

\bibitem{Murakami1997} Murakami, A., Sakaguchi, H., Hasegawa, T.: Dislocation, vortex and couple stress in the formation of shear bands under trap-door problems. Soils and found. 37, 123-135 (1997). https://doi.org/10.3208/sandf.37.123

\bibitem{Zhang2013} Zhang, W., Wang, J., Jiang, M.: DEM-Aided Discovery of the Relationship between Energy Dissipation and Shear Band Formation Considering the Effects of Particle Rolling Resistance. J. of Geotech. and Geoenvironmental Eng. 139, 1512-1527 (2013). https://doi.org/10.1061/(asce)gt.1943-5606.0000890

\bibitem{Tang2016} Tang, H., Dong, Y., Chu, X., Zhang, X.: The influence of particle rolling and imperfections on the formation of shear bands in granular material. Granul. Matter 18, 1-12 (2016). https://doi.org/10.1007/s10035-016-0607-3

\bibitem{Nho2021} Nho, H., Nguyen, G., Scholt{\`{e}}s, L., Guglielmi, Y., Victor, F.: Micromechanics of sheared granular layers activated by fluid pressurization Micromechanics of sheared granular layers activated by fluid pressurization. Geophys. Res. Lett., 48, e2021GL093222 (2021). https://doi.org/10.1002/essoar.10506504.1

\bibitem{Estrada2008} Estrada, N., Taboada, A., Radja{\"{i}}, F.: Shear strength and force transmission in granular media with rolling resistance. Phys. Rev. E - Stat., Nonlinear, and Soft Matter Phys. 78, 1-11 (2008). https://doi.org/10.1103/PhysRevE.78.021301

\bibitem{Yang2017} Yang, Y., Cheng, Y.M., Sun, Q.C.: The effects of rolling resistance and non-convex particle on the mechanics of the undrained granular assembles in 2D. Powder Technol. 318, 528-542 (2017). https://doi.org/10.1016/j.powtec.2017.06.027

\bibitem{Liu2018} Liu, Y., Liu, H., Mao, H.: The influence of rolling resistance on the stress-dilatancy and fabric anisotropy of granular materials. Granul. Matter 20, 12 (2018). https://doi.org/10.1007/s10035-017-0780-z

\bibitem{Barnett2020} Barnett, N., Mizanur Rahman, Md., Rajibul Karim, Md., Nguyen, H.B.K.: Evaluating the particle rolling effect on the characteristic features of granular material under the critical state soil mechanics framework. Granul. Matter 22, 89 (2020). https://doi.org/10.1007/s10035-020-01055-5

\bibitem{GODET1984437} Godet, M.: The third-body approach: A mechanical view of wear. Wear 100, 437-452 (1984). https://doi.org/10.1016/0043-1648(84)90025-5

\bibitem{COLAS2013192} Colas, G., Saulot, A., Godeau, C., Michel, Y., Berthier, Y.: Decrypting third body flows to solve dry lubrication issue – MoS2 case study under ultrahigh vacuum. Wear 305, 192-204 (2013). https://doi.org/10.1016/j.wear.2013.06.007

\bibitem{https://doi.org/10.1002/(SICI)1096-9853(199905)23:6<531::AID-NAG980>3.0.CO;2-V} Jensen, R.P., Bosscher, P.J., Plesha, M.E., Edil, T.B.: DEM simulation of granular media—structure interface: effects of surface roughness and particle shape. Int. J. for Numer. and Anal. Methods in Geomech. 23, 531-547 (1999). https://doi.org/10.1002/(SICI)1096-9853(199905)23:6<531::AID-NAG980>3.0.CO;2-V

\bibitem{Kozicki2011} Kozicki, J., Tejchman, J.: Numerical simulations of sand behavior using DEM with two different descriptions of grain roughness. In: Oñate, E., Owen, D.R.J. (Eds) II International Conference on Particle-based Methods – Fundamentals and Applications. Particles 2011 (2011)

\bibitem{Mollon2020} Mollon, G., Quacquarelli, A., And{\`{o}}, E., Viggiani, G.: Can friction replace roughness in the numerical simulation of granular materials ?. Granul. Matter 22, 42 (2020). https//doi.org/10.1007/s10035-020-1004-5

\bibitem{Garcia2009} Garcia, X., Latham, J.P., Xiang, J., Harrison, J.: A clustered overlapping sphere algorithm to represent real particles in discrete element modelling. Geotech. 59, 779-784 (2009). https://doi.org/10.1680/geot.8.T.037

\bibitem{Podlozhnyuk2018} Podlozhnyuk, A.: Modelling superquadric particles in DEM and CFD-DEM: implementation, validation and application in an open-source framework. (2018)

\bibitem{Cundall1988} Cundall, P.: Formulation of a three-dimensional distinct element model-Part I. A scheme to detect and represent contacts in a system composed of many polyhedral blocks. Int. J. Rock Mech. Min. Sci. \& Geomech 25, 107-116 (1988). https://doi.org/10.1016/0148-9062(88)92293-0

\bibitem{Nezami2004} Nezami, E.G., Hashash, Y.M.A., Zhao, D., Ghaboussi, J.: A fast contact detection algorithm for 3-D discrete element method. Comput. and Geotech. 31, 575-587 (2004). https://doi.org/10.1016/j.compgeo.2004.08.002

\bibitem{AlonsoMarroquin2009} Alonso-Marroquin, F., Wang, Y.: An efficient algorithm for granular dynamics simulations with complex-shaped objects. Granul. Matter 11, 317-329 (2009). https://doi.org/10.1007/s10035-009-0139-1

\bibitem{Wensrich2012} Wensrich, C.M., Katterfeld, A.: Rolling friction as a technique for modelling particle shape in DEM. Powder Technol. 217, 409-417 (2012). https://doi.org/10.1016/j.powtec.2011.10.057

\bibitem{Rorato2021} Rorato, R., Arroyo, M., Gens, A., And{\`{o}}, E., Viggiani, G.: Image-based calibration of rolling resistance in discrete element models of sand. Comput. and Geotech. 131, 103929 (2021) https://doi.org/10.1016/j.compgeo.2020.103929

\bibitem{Jiang2005} Jiang, M.J., Yu, H.S., Harris, D.: A novel discrete model for granular material incorporating rolling resistance. Comput. and Geotech. 32, 340-357 (2005). https://doi.org/10.1016/j.compgeo.2005.05.001

\bibitem{Johnson1985} Johnson, K.L.: Contact Mechanics. Cambridge University Press, London (1985). https://doi.org/10.1017/CBO9781139171731

\bibitem{Kloss2012} Kloss, C., Goniva, C., Hager, A., Amberger, S., Pirker, S.: Models, algorithms and validation for opensource DEM and CFD-DEM. Prog. in Comput. Fluid Dyn. 12, 140-152 (2012). https://doi.org/10.1504/PCFD.2012.047457

\bibitem{Marone2005} Anthony, J., Marone, C.: Influence of particle characteristics on granular friction. J. of Geophys. Res.: Solid Earth 110, 1-14 (2005). https://doi.org/10.1029/2004JB003399

\bibitem{Koval2011} Koval, G., Chevoir, F., Roux, J.N., Sulem, J., Corfdir, A.: Interface roughness effect on slow cyclic annular shear of granular materials. Granul. Matter 13, 525-540 (2011). https://doi.org/10.1007/s10035-011-0267-2

\bibitem{Rattez2020} Rattez, H., Shi, Y., Sac--Morane, A., Klaeyle, T., Mielniczuk, B., Veveakis, M.: Effect of grain size distribution on the shear band thickness evolution in sand. Géotech. 72, 350-363 (2020). https://doi.org/10.1680/jgeot.20.P.120

\bibitem{Dieterich1979} Dieterich, J.H.: Modeling of rock friction 1. Experimental results and constitutive equations. J. of Geophys. Res.: Solid Earth 84, 2161-2168 (1979). https://doi.org/10.1029/JB084iB05p02161

\bibitem{Morrow1989} Morrow, C.A., Byerlee, J.D.: Experimental studies of compaction and dilatancy during frictional sliding on faults containing gouge. J. of Struct. Geol. 11, 815-825 (1989). https://doi.org/10.1016/0191-8141(89)90100-4

\bibitem{Ferdowsi2020} Ferdowsi, B., Rubin, A.M.: A Granular Physics-Based View of Fault Friction Experiments. J. of Geophys. Res.: Solid Earth 125, 1-32 (2020). https://doi.org/10.1029/2019JB019016

\bibitem{Beroza1990} Beroza, G.C., Jordan, T.H.: Searching for slow and silent earthquakes using free oscillations. J. of Geophys. Res. 95, 2485-2510 (1990). https://doi.org/10.1029/JB095iB03p02485 

\bibitem{THORNTON1988133} Thornton, C., Randall, C.W.: Applications of Theoretical Contact Mechanics to Solid Particle System Simulation. Micromech. of Granul. Mater. 20, 133-142 (1988). https://doi.org/10.1016/B978-0-444-70523-5.50023-0

\bibitem{Li2005} Li, Y., Xu, Y., Thornton, C.: A comparison of discrete element simulations and experiments for 'sandpiles' composed of spherical particles. Powder Technol. 160, 219-228 (2005). https://doi.org/10.1016/j.powtec.2005.09.002

\bibitem{Roux2002} Roux, J.N., Combe, G.: Quasistatic rheology and the origins of strain. C. R. Phys. 3, 131-140 (2002). https://doi.org/10.1016/S1631-0705(02)01306-3

\bibitem{DaCruz2005} Da Cruz, F., Emam, S., Prochnow, M., Roux, J.N., Chevoir, F.: Rheophysics of dense granular materials: Discrete simulation of plane shear flows. Phys. Rev. E - Stat., Nonlinear, and Soft Matter Phys. 72, 1-17 (2005). https://doi.org/10.1103/PhysRevE.72.021309

\bibitem{Roux2010} Roux, J.N., Chevoir, F.: Analyse dimensionnelle et param{\`{e}}tres de contr{\^{o}}le. In: Radja{\"{i}}, F., Dubois, F. (eds.) Mod{\'{e}}lisation num{\'{e}}rique discr{\`{e}}te des mat{\'{e}}riaux granulaires, pp 223-259. Herm{\`{e}}s - L, Paris (2010)

\bibitem{Jaeger1996} Jaeger, H.M., Nagel, S.R., Behringer, R.P.: Granular solids, liquids, and gases. Rev. Mod. Phys. 68, 1259-1273 (1996). https://doi.org/10.1103/RevModPhys.68.1259

\bibitem{Zhu2021} Zhu, F., Zhao, J.: Interplays between particle shape and particle breakage in confined continuous crushing of granular media. Powder Technol. 378, 455-467 (2021). https://doi.org/10.1016/j.powtec.2020.10.020

\bibitem{Ueda2013} Ueda, T., Matsushima, T., Yamada, Y.: DEM simulation on the one-dimensional compression behavior of various shaped crushable granular materials. Granul. Matter 15, 675-684 (2013). https://doi.org/10.1007/s10035-013-0415-y

\bibitem{Zhang2018} Zhang, X., Hu, W., Scaringin G., Baudet, B.A., Han, W.: Particle shape factors and fractal dimension after large shear strains in carbonate sand. Geotech. Lett. 8, 73-79 (2018). https://doi.org/10.1680/jgele.17.00150

\bibitem{Buscarnera2021} Buscarnera, G., Einav, I.: The mechanics of brittle granular materials with coevolving grain size and shape. Proc. R. Soc. A 477, 20201005 (2021). https://doi.org/10.1098/rspa.2020.1005

\bibitem{Morgan1999} Morgan, J.K.: Numerical simulations of granular shear zones using the distinct element method: 2. Effects of particle size distribution and interparticle friction on mechanical behavior. J. of Geophys. Res.: Solid Earth 104, 2721-2732 (1999). https://doi.org/10.1029/1998jb900055

\bibitem{Sammis1987} Sammis, C., King, G., Biegel, R.: The kinematics of gouge deformation. Pure and Appl. Geophys. 125, 777–812 (1987). https://doi.org/10.1007/BF00878033

\bibitem{Einav2007} Einav, I.: Breakage mechanics—part i: Theory. J. of the Mech. and Phys. of Solids 55, 1274–1297 (2007). https://doi.org/10.1016/j.jmps.2006.11.003

\bibitem{Rattez2021} Rattez, H., Disidoro, F., Sulem, J., Veveakis, M.: Influence of dissolution on long-term frictional properties of carbonate fault gouge. Geomech. for Energy and the Environment 26, 100234 (2021). https://doi.org/10.1016/j.gete.2021.100234.

\end{thebibliography}
\end{document}